# High-Frequency Tunable Resistorless Memcapacitor Emulator and Application

Pratik Kumar[†], Sajal K. Paul[*]

pratikkumar@iisc.ac.in, sajalkpaul@rediffmail.com

[†]Department of Electronic Systems Engineering, Indian Institute of Science, Bangalore, India, 560012

[†]Centre for Nano Science and Engineering, Indian Institute of Science, Bangalore, India, 560012

*Department of Electronics Engineering, Indian Institute of Technology (ISM), Dhanbad, India, 860024

## Abstract

In this paper, a new design has been proposed for the realization of high-frequency memcapacitor emulators built with three OTAs. This paper also proposes the application of memcapacitor as an amplitude modulator. Furthermore, applications of memcapacitor as a filter, Oscillator point attractor, and periodic doubler are also shown. The proposed circuits can be configured in both incremental and decremental topology. The proposed circuits and their application claim that the circuit is much simpler in design and can be utilized in both topologies. The performance of all the proposed circuits has been verified on Cadence Virtuoso Spectre using standard CMOS 180nm. Furthermore, post-layout simulations and their comparison have been carried out.

## 1. Introduction

Apart from resistor, inductor, and capacitor forming the three fundamental basic electrical elements, memristor (MR) now represents the fourth fundamental element. Memristor was postulated by Chua in 1971 [1] as the fourth basic electrical element but came into highlight only often HP in 2008 fabricated a memristor based on thin film TiO2. From then onward there is a boom of research in this field. In 1980 [2] this postulation was then generalized to an infinite variety of basic circuit elements generally called element quadrangle. Chua's circuit element quadrangle was then extended to propose higher order elements such as memcapacitor(MC) and meminductor (MI). The memcapacitance of the memcapacitor represents links between the magnetic flux $\phi$ and the time integral of the charge $\sigma$, while the meminductance provides a relationship between the charge q and the time integral of flux $\rho$. Unlike capacitor and inductor, they can store information for a very long time without power because of their non-volatility. Although the device is still a theoretical concept, but emulators are required to analyze its characteristics and study its applications. In 2009 [3] memristive system was extended to capacitive and inductive elements whose properties depend on the state and history of the system [3]. A physical and characteristic analysis of these memory based elements along with mathematical examples for memristor, meminductor, and memcapacitor were presented in [4-5].

Several circuits for emulating MC has been also proposed in [6-16] by taking advantage of active devices and memristors. [6] proposed the memcapacitor emulator using memristor, consisting of a potentiometer, microcontroller, ADC and other passive components along with op-Amp. Mutator simulating second order elements and its relationship as well as the implementation of MC←MR transformation is shown in [7]. Mutator for transforming memristor into memcapacitor consisting of active and passive devices is shown in [8]. Realization of emulators for transforming memristor device into effective floating memcapacitor and meminductor system based on second generation Current Conveyor along with passive components as proposed in [9]. [10] shows the charge controlled first memristor less memcapacitor flux controlled floating emulator and its mathematical model. Design of memcapacitor based on LDR memristor was shown in [11]. Memcapacitor emulator design using memristor was proposed in [12]. MC emulators in serial, parallel, hybrid, and even complicated Wye and Delta connections with multiple input sources are proposed on the basis of an expandable memristor emulator along with its mathematical analysis is proposed in [13]. However, most of these earlier reported emulators have the restriction of the frequency of operation to few kHz and a larger number of active as well as passive components. Later on in [14] a universal mutator for transformation among three memory elements using three Transimpedance Operational Amplifier (TOAs) was proposed. [15] showed a unique approach for a memcapacitor based on stacking a memristor with a combination of traditional MIM. Capacitors and memristor a method for emulating floating MCs with piecewise-linear constitutive relations between voltage and current is presented based on the multiple-state floating capacitor in [16]. Circuits proposed in [6-16] are operating only at low frequency range and/or uses a large number of active and passive components.

[17-18] proposes the application of memcapacitor as couples and an uncoupled relaxation oscillator. [19] shows the small signal model of memcapacitor and its memcapacitor-inductor based oscillator. [20-21] shows the LPF design using memristor, memcapacitor, and meminductor.

In [23] shows that Pt-Fe203 core- shell nanoparticles assembly on P+-Si substrate exhibits analog memristive and memcapacitive characteristics. [23-26] shows that the memcapacitor can be used to perform neuromorphic, digital, quantum calculation, data storage and synapses. Application of memcapacitor as chaotic oscillator was further proposed and studied in [27-31]. [32] also shows that existing co-attractors were present in memcapacitor based chaotic circuits.

The paper proposes high frequency memcapacitor emulator built with only three OTA and two grounded capacitors. The proposed memcapacitor emulators possess following important features : (i) simple circuitry, (ii) option for both incremental and decremental configurations to increase the range of values of memcapacitance for application flexibility [24], (iii) high frequency range of operation, (iv) resistor less emulator, (v) electronic control of memcapacitance value in addition to the control by frequency and amplitude of the applied voltage signal across emulator and (vi) high input impedance for voltage input.

## 2. Operational transconductance amplifier (OTA) circuit

Circuit symbol of Operational Transconductance Amplifier (OTA) is shown in Fig.1, in which there the availability of high input impedance terminals for voltage and a high impedance output terminal for current along with electronically tunable transconductance gain ($G_m$) and Bias current.

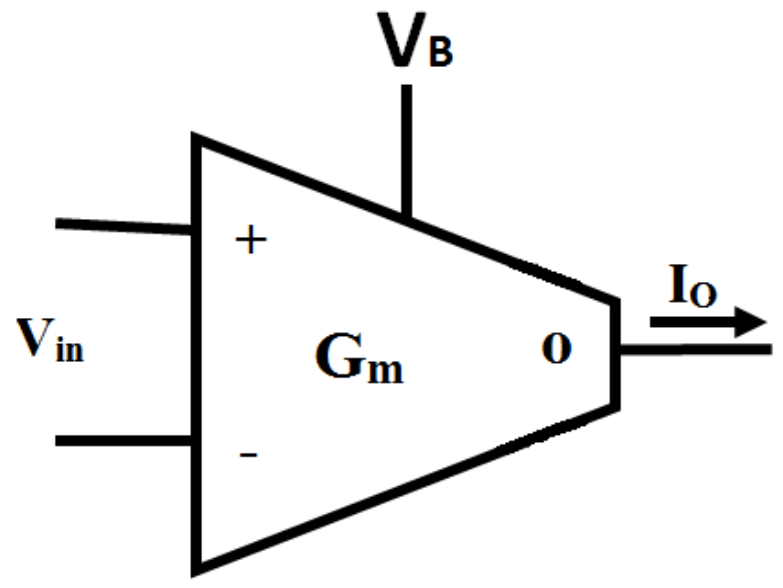

(a)

**Fig. 1.** Symbolic representation of OTA for (a) single output

The CMOS implementation of OTA is given in Fig.2. The output of an OTA for an input $V_{in}$ is expressed as

$$I_{O\pm} = \pm G_m V_{in} \quad , \quad V_{in+} - V_{in-} = \text{differential input} \tag{1}$$

Where $G_m$ is transconductance of OTA. The routine analysis results in the expression of $G_m$ as

$$G_m = \frac{k}{\sqrt{2}}(V_B - V_{ss} - 2V_{th}) \,, \tag{2}$$

where k is a parameter of MOS device given by

$$k = \mu_n C_{ox} \frac{W}{L} \tag{3}$$

The W, L, $\mu_n$, $C_{ox}$ and $V_{th}$ are respectively channel width, channel length, the mobility of the carrier, capacitance per unit area and the threshold voltage of MOS. Fig. 3 shows the frequency response of the OTA of Fig. 2.

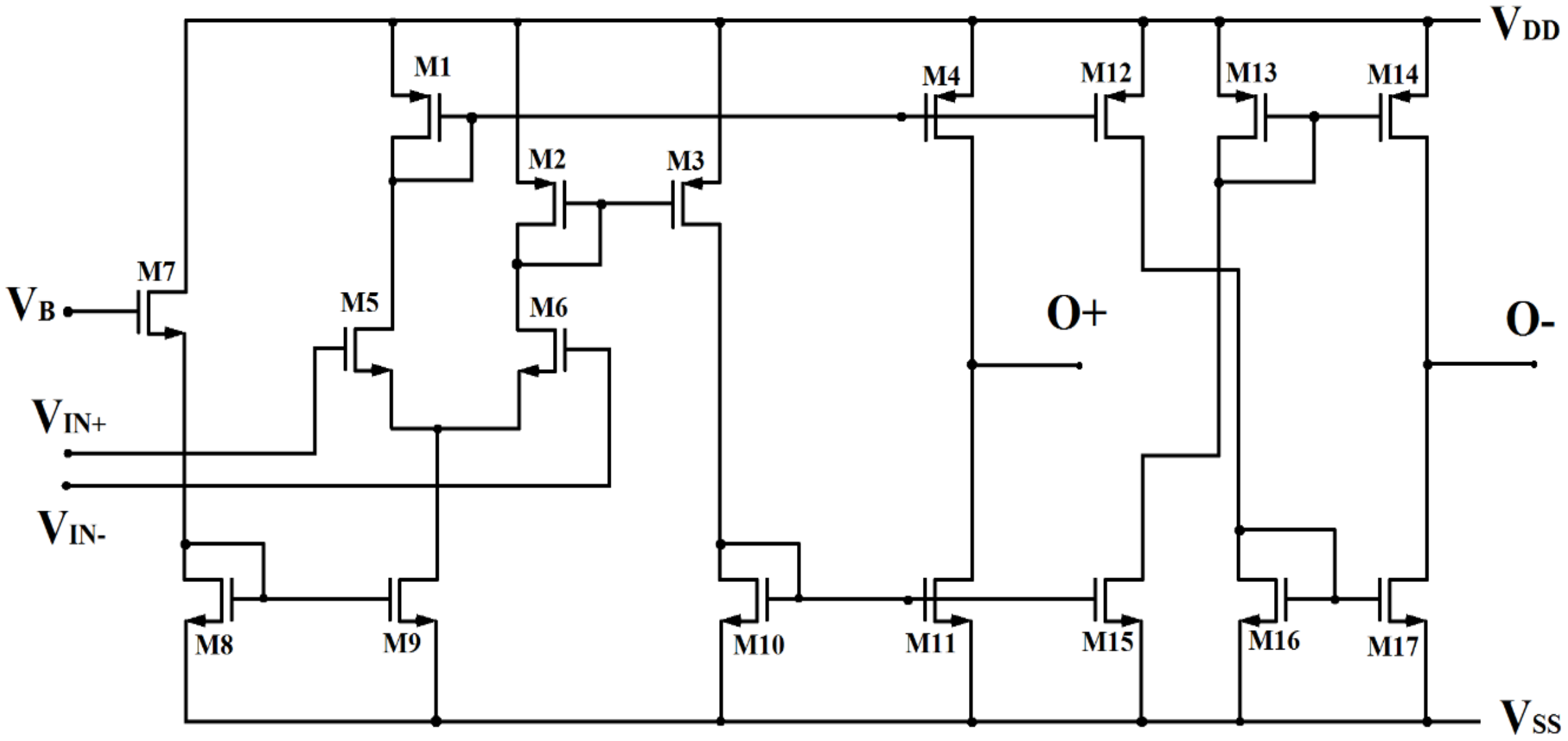

**Fig. 2.** CMOS single Output OTA circuit

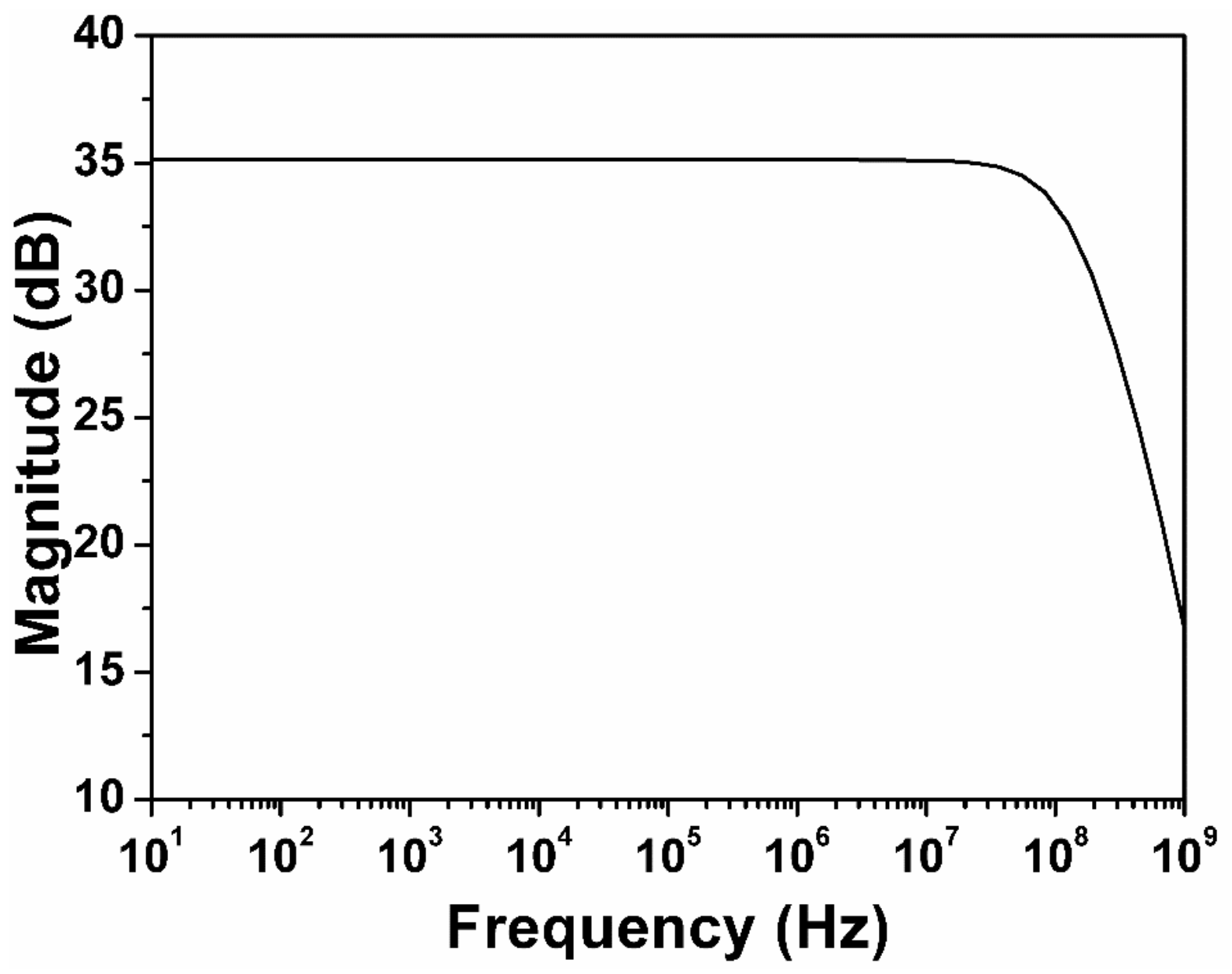

**Fig. 3.** Frequency response of OTA used for memristor emulator.

## 3. Proposed memcapacitor emulator circuits

Schematic diagrams of the proposed memcapacitor emulator are shown in Fig.4. The Incremental and decremental nature of memcapacitor can be configured by switching mechanism among pins w, x, y and z of circuits as given in Table 1.

**Table 1.** Connection topology for pins w, x, y and z for two modes of operations.

| S. No. | Switch Connections | Mode of operation |
|---|---|---|
| 1 | w-x; y-z | Incremental |
| 2 | w-z; y-x | Decremental |

The proposed memcapacitor is shown in Fig. 4. Considering incremental type memcapacitor emulator i.e., pins w, x and pins y, z are interconnected, the input current $I_{in}$ and capacitor current $I_C$ are obtained as

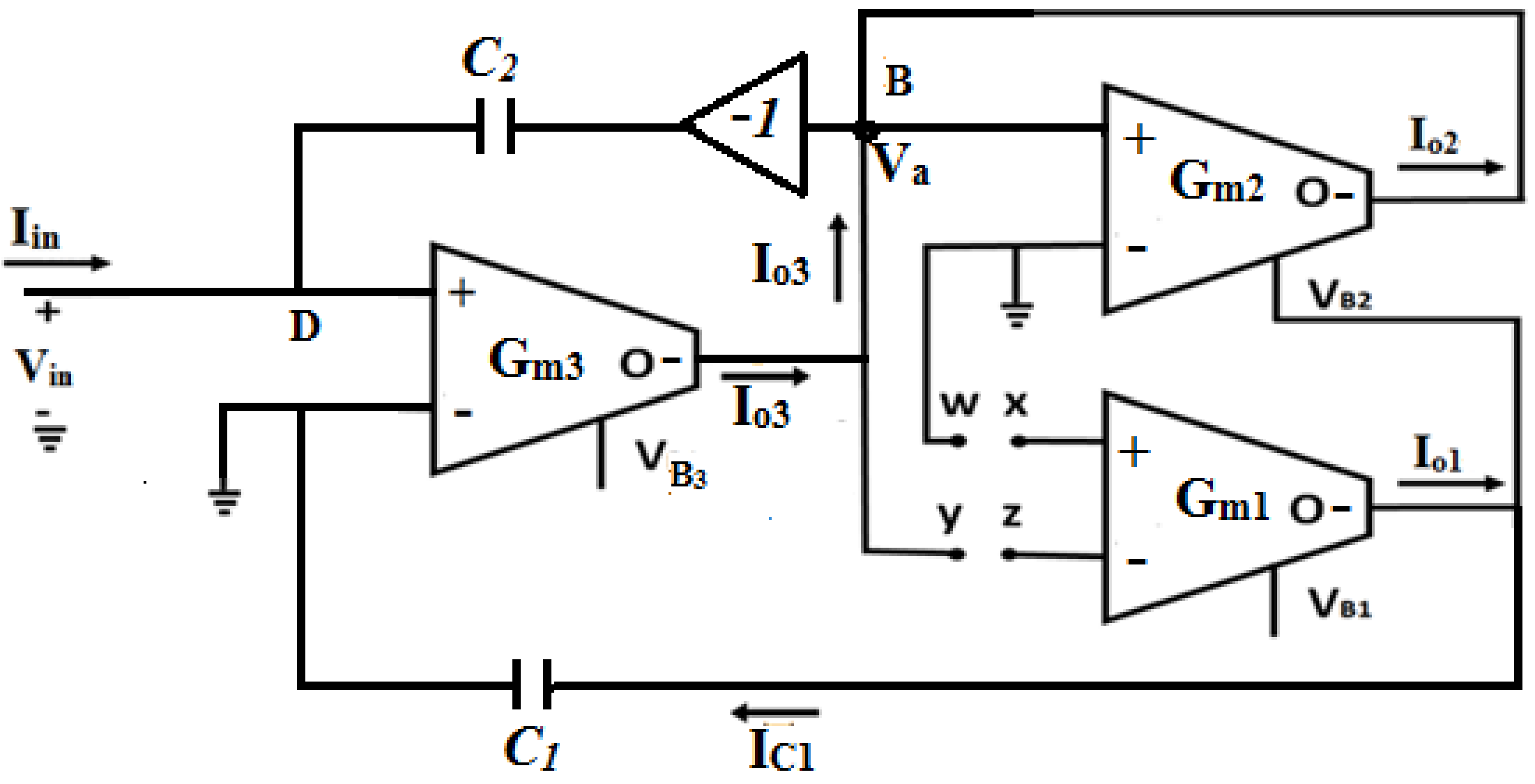

**Fig. 4.** Schematic diagram of memcapacitor emulator

For w-x , y-z

$$I_{02} = -G_{m2}V_a(t) = -I_{03} = G_{m3}V_{in}(t) \tag{4}$$

$$I_{C1}(t) = I_{01} = G_{m1}V_a(t) \tag{5}$$

Bias voltage $V_{B2}$ is given by

$$V_{B2} = \frac{1}{C_1}\int I_{C1}(t)dt = \frac{G_{m1}}{C_1}\int V_a(t)dt \tag{6}$$

Substituting (6) into (2), transconductance $G_{m2}$ is obtained as

$$G_{m2} = \frac{k}{\sqrt{2}}\left(V_{B2} - V_{ss} - 2V_{th}\right) = \frac{k}{\sqrt{2}}\left(\frac{G_{m1}}{C_1}\int V_a(t)dt - V_{ss} - 2V_{th}\right) \tag{7}$$

Combining (4) and (8) results in an expression for input current as

$$I_{O3} = \frac{k}{\sqrt{2}}\left(\frac{G_{m1}}{C_1}\int V_a(t)dt - V_{ss} - 2V_{th}\right)V_a \tag{8}$$

Hence the memcapacitance of the proposed incremental memcapacitor emulator is obtained as

$$I_{o3} = \frac{V_a}{\beta_1} \tag{9}$$

Where

$$\frac{1}{\beta 1} = \frac{k}{\sqrt{2}}\left(\frac{G_{m1}}{C_1}\int V_a(t)dt - V_{ss} - 2V_{th}\right)$$

From (4) and (9)

$$V_a = -G_{m3}V_{in}(t)\beta_1 \tag{10}$$

also, $$I_{in} = (V_{in} + V_a)\mathrm{SC}_3 \tag{11}$$

Substituting (10) in (11) and using the value of $\beta_1$ we get

$$\frac{\mathrm{V}_{in}}{I_{in}} = \frac{1}{S\,\mathrm{C}_3\left[1 + \dfrac{G_{\mathrm{m}3}}{\dfrac{k}{\sqrt{2}}\left(\dfrac{G_{m1}}{C_1}\int V_a(t)dt + V_{ss} + 2V_{th}\right)}\right]} \tag{12}$$

On applying KCL at node D

$$V_a = -\frac{G_{m3}}{G_{m2}} V_{in} \tag{13}$$

Using (12) and (13) we get

$$\frac{V_{in}}{I_{in}} = \frac{1}{S\,C_3\left[1+\dfrac{G_{m3}}{\dfrac{k}{\sqrt{2}}\left(V_{ss}+2V_{th}-\dfrac{G_{m1}G_{m3}}{G_{m2}C_1}\int V_{in}(t)dt\right)}\right]} \tag{14}$$

Where the memcapacitance $C_M$ is

$$(C_m) = C_3\left[1+\frac{Gm_3}{\dfrac{k}{\sqrt{2}}\left(V_{ss}+2V_{th}-\dfrac{Gm_1Gm_3}{Gm_2C_1}\int V_{in}dt\right)}\right] \tag{15}$$

Similarly, the change of switch connections to w-z and y-x changes polarity of time variant part of memcapacitor of (16), resulting in a decremental type memcapacitance as

$$\frac{V_{in}}{I_{in}} = \frac{1}{S\,C_3\left[1+\dfrac{G_{m3}}{\dfrac{k}{\sqrt{2}}\left(V_{ss}+2V_{th}+\dfrac{G_{m1}G_{m3}}{G_{m2}C_1}\int V_{in}(t)dt\right)}\right]} \tag{16}$$

In equation (15) and (16) $G_{m1}$ and $G_{m3}$ can be controllable by external bias voltages $V_{B1}$ and $V_{B3}$ respectively, which makes proposed circuit electronically tunable. The equations (15) and (16) represent incremental and decremental memcapacitance respectively, where $S_{C3}$ is the constant term for a fixed frequency and $\left[\dfrac{S\,C_3\,G_{m3}}{\dfrac{k}{\sqrt{2}}\left(V_{ss}+2V_{th}\mp\dfrac{G_{m1}G_{m3}\phi_{in}}{G_{m2}C_1}\right)}\right]$ being the time-varying term as $\phi_{in}$ is the function of time varying input signal. For $\phi_{in}=0$ the memcapacitance attains a constant capacitance value in both the topology (incremental and decremental). Furthermore, for sinusoidal input signal $V_{in}(t)=A_m \, Sin(\omega t)$, the $\phi_{in}(t)$ results as

$$\phi_{in}(t)=\frac{A_m}{\omega}\cos\left(\omega t-\frac{\pi}{2}\right) \tag{17}$$

where $A_m$ is the amplitude of the voltage signal applied and ω is the frequency of the signal in radian per second which on substituting in (15) and (16) gives

$$\frac{I_{in}}{V_{in}}=S\,C_3\left[1+\frac{G_{m3}}{\dfrac{k}{\sqrt{2}}\left(V_{ss}+2V_{th}\pm\dfrac{G_{m1}G_{m3}\varphi_{in}(t)\,A_m\cos\left(\omega t-\frac{\pi}{2}\right)}{G_{m2}C_1\omega}\right)}\right] \tag{18}$$

where for the operator ±, the + is for incremental and – is for decremental configuration. For all the further derivations and simulations $C_2$ is assumed to be a constant capacitor. $C_2$ can be varied from pf to nf as it acts only as an integrator and does not affect time varying part of memcapacitor directly.

## 3.1. Comparison of memcapacitor emulators

Comparison of available memcapacitor emulators and mutators for MC ← MR are given in Table 2. It can be observed that most of the emulators and mutators use complex analog building blocks along with a large number of passive components for implementation of memcapacitor emulators, which is not normally encouraged in integrated circuits. In comparison with the proposed circuit, it can be verified that the proposed circuits are simple and use only three OTAs.

The CCII based emulators in [5] use memristor based on a microcontroller, ADC, potentiometer as well as a resistor, capacitor and opAmp. [6] proposes mutators using 1 CCII, capacitor, and 1 OTA. [7] proposes mutators using 2 CCII and 1 resistor. [8] is however floating but utilizes a large number of active and passive components such as 4 CCII, inductor, resistor and memristor for memcapacitor realization. [10] utilizes a simple method to realize charge controlled memcapacitor but uses a multiplier, mirror, capacitance resistances, 2 OTA. [11-12] uses a very large number of active and passive components. These do not have electronic tunability as well. Moreover, the frequency range is few Hz to few KHz only. Whereas, the proposed emulator is useful respectively for 8 MHz. They are resistorless and uses only two capacitors and hence suitable for IC implementation. Both the incremental and decremental properties are present in the proposed emulators. An important feature of the proposed memcapacitor is its ability to control the memcapacitive value by controlling the transconductance $G_{m1}$ and $G_{m3}$ by bias voltage $V_{B1}$ and $V_{B3}$.

**Table 2.** Comparison of memcapacitor emulators.

*Denotes potentiometer is used as variable resistance

| **Ref.** | **Number in count and type(s) of active building blocks used** | **Passive Elements ($L/R/C/R_M$)** | **Resistor-less** | **Electronic Tunability** | **Tech. Used** | **Max. Frequency of operation** |
|---|---|---|---|---|---|---|
| [6] | 1 ADC, 1 uc,1 Op-Amp | 0/2*/1/1 | No | No | CMOS | Few Hz |
| [7] | 1 OTA, 1 CCII | 0/0/1/1 | No | No | CMOS | Few Hz |
| [8] | 2 CFOAs | 0/1/1/1 | No | No | CMOS | Few Hz |
| [9] | 4 CFOAs | 1/1/0/1 | No | No | CMOS | Few Hz |
| [10] | 2 Op-Amp, 1 Current Mirror,1 Buffer, 1 Multiplier | 0/2/3/0 | No | No | CMOS | Few Hz |
| [11] | 2 CFOAs, 1 diode,3 Op-Amp | 0/14/2/1 | No | No | CMOS | Few Hz |
| [12] | 4 CCII | 0/1/2/1 | No | No | CMOS | Few KHz |
| [13] | 2 CCII, 2 Buffer | 0/1/1/1 | No | No | CMOS | Few Hz |
| [14] | 3 CCII, 3 Buffer | 0/3/1/1 | No | No | CMOS | 1 MHz |
| [16] | 4 Op-Amp | 0/6*/4/1 | No | No | CMOS | Few Khz |
| Proposed Work | 3 OTA, 1 Buffer | 2/0 | Yes | Yes | CMOS | 8 MHz |

# 4. Simulations results and discussion

In this section memcapacitive properties (hysteresis loop and non-volatility) of the proposed circuits have been verified. To verify memcapacitive nature of proposed emulator circuits, simulation with 180 nm TSMC technology has been performed. A supply voltage of +1.2V and -1.2 V for $V_{DD}$ and $V_{SS}$ respectively for the OTA circuit shown in Fig. 2 is used. All the MOS transistors are operated in the saturation region.

## 5.1. *Memcapacitance obtained from charge vs. time plot*

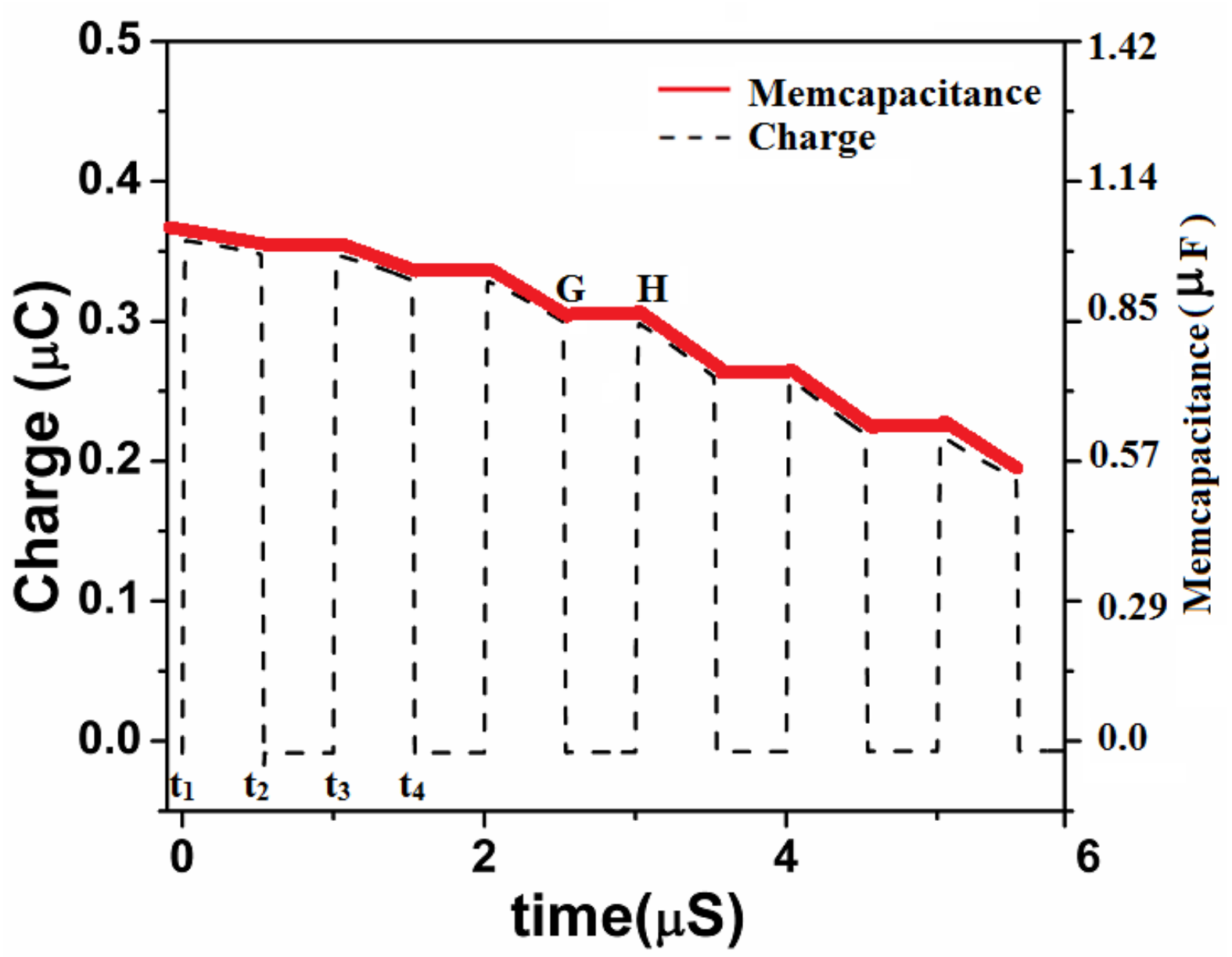

**Fig. 5.** Charge vs. time and memcapacitance vs. time for 350mV pulse input for a time period of 100ns at an input frequency of 1 MHz, $V_{B1,3}$ = 0.5V, and Capacitance =220 pf for decremental topology.

A 350mV input voltage pulse with definite pulse width is applied as an input to the memcapacitor emulator circuit of Fig. 5. The simulated memcapacitance vs. time graph is shown in Fig. 5 by dark lines for decremental topology. The end peak of one pulse is joined with the starting peak of next pulse (for e.g. point G-H) which comes out to be a straight line. This straight line when combined with the slope line of the charge vs. time plot divided by some scaling factor (here 350mv) results in memcapacitance vs. time as shown in Fig. 5.

From Fig. 5, it can be observed that memcapacitance vs. time plot is decreasing in the interval t1 - t2. For the next ON pulse interval t3-t4, the previous value of memcapacitance at t2 is retained and it again decreases from the previously retained value for decremental topology. This is confirmed by each successive pulse period.

The interval t1-t2 or each ON pulse interval corresponds to the time duration where memcapacitance decreases for decremental topology. This property of decreasing memcapacitance is similar to memcapacitor property where on the application of input signal memcapacitance changes its state or value. Similarly, the interval t2-t3 or each pulse OFF period can be said as an interval where memcapacitance nature of the memcapacitor is conserved. This property of constant memcapacitance is similar to non-volatility property where previously attained value is maintained by the memcapacitor even after withdrawal of pulse or signal.

### *5.2. Memcapacitor simulation result*

The memcapacitor of Fig. 4 is simulated for different frequencies with aspect ratio, as given in Table 3, for MOS transistors of single output OTA. The results for pinched hysteresis loop obtained for frequencies of 500 Hz,1 KHz, 500 KHz, 1MHz, 5MHz, 4 MHz, and 8 MHz are shown in Fig. 6. Here the product of capacitor(C) value and frequency (f) is kept constant (0.28 FaradHz) with $A_m$=350mV, $V_{B1}$=640mV. The overlapping of pinched hysteresis nature of q-v curves validates the memcapacitive behavior of emulators as obtained in (15).

Fig. 7 shows the single-valued relationship between C(t) and $\Phi$(t). The locus of  shows the single-valued curved with respect to $\Phi$.Therefore the corresponding device is a memcapacitor.

The response of the memcapacitive emulator circuits is also obtained for a constant voltage amplitude pulse with a definite rise($t_r$) and fall time($t_f$) of 15ns and a definite pulse period ($t_p$) as 1us for the proposed memcapacitive emulator. The bias voltage is maintained at 500 mV while capacitance is varied to maintain the product of Cf at 2x10$^{-4}$ FaradHz. As discussed in Section 5.1 by simply looking at the current versus time plot one can judge the memcapacitance if a voltage pulse is applied at the input. An analysis of Fig 8(a-b) reveals that the memcapacitance started from a constant value and then increasing or decreasing for incremental and decremental topology respectively which is in agreement with (15) and (16).

**Table 3. Design Parameters of single output OTA of Fig. 2 for memcapacitor**

| MOS Transistors | W(μm) | L(nm) |
|---|---|---|
| $M_{1-4}$ | 12 | 375 |
| $M_{10}$ | 12 | 510 |
| $M_{5-9}$, $M_{11}$ | 12 | 500 |

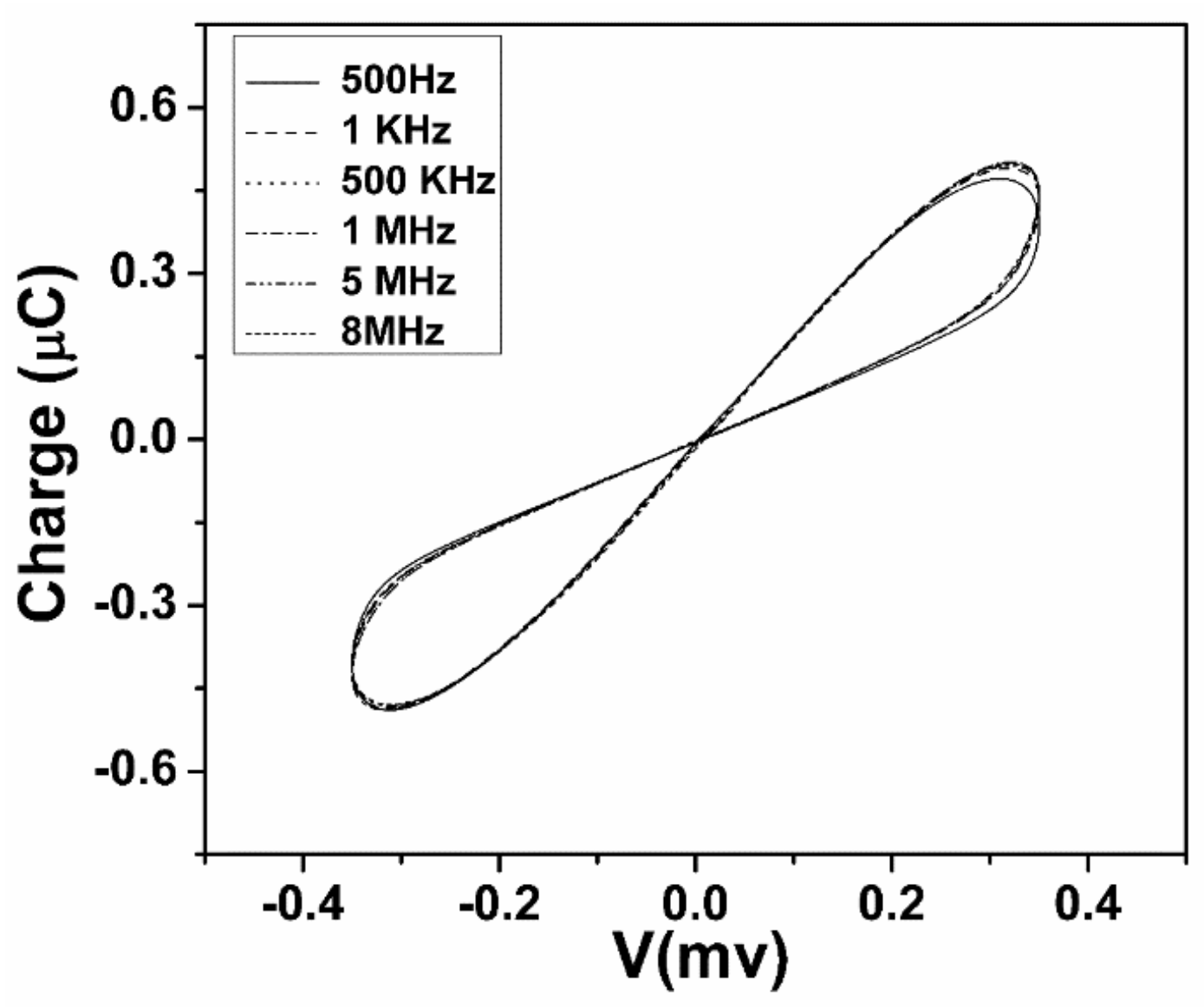

**Fig. 6.** q-v characteristic for memcapacitor circuit at different operating frequencies for $V_{B1}$= 0.6 V, Vin = 450 mV, and constant Cf =2*10$^{-4}$FaradHz

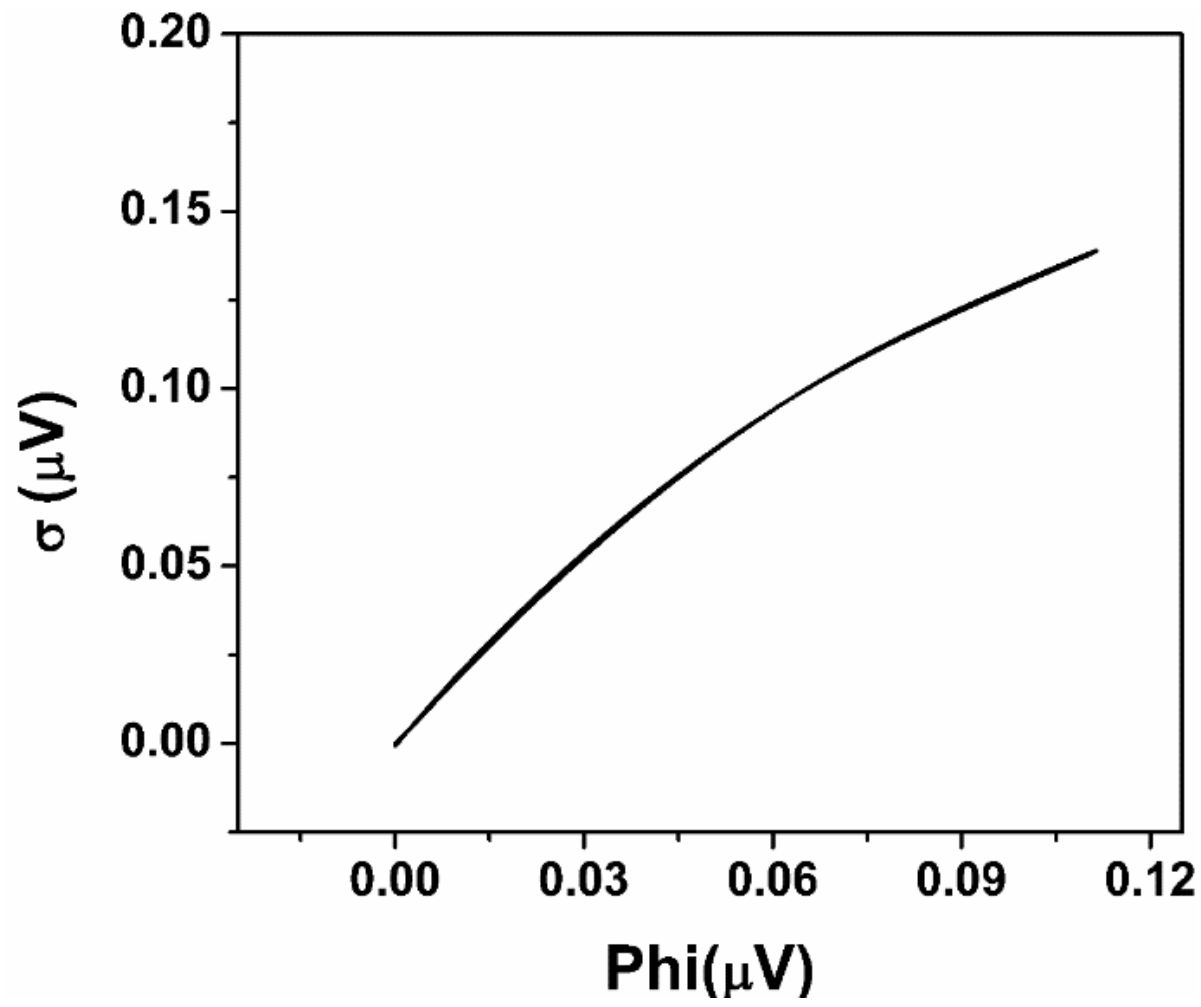

**Fig. 7.** Locus of Φ verses $\sigma$ characteristics

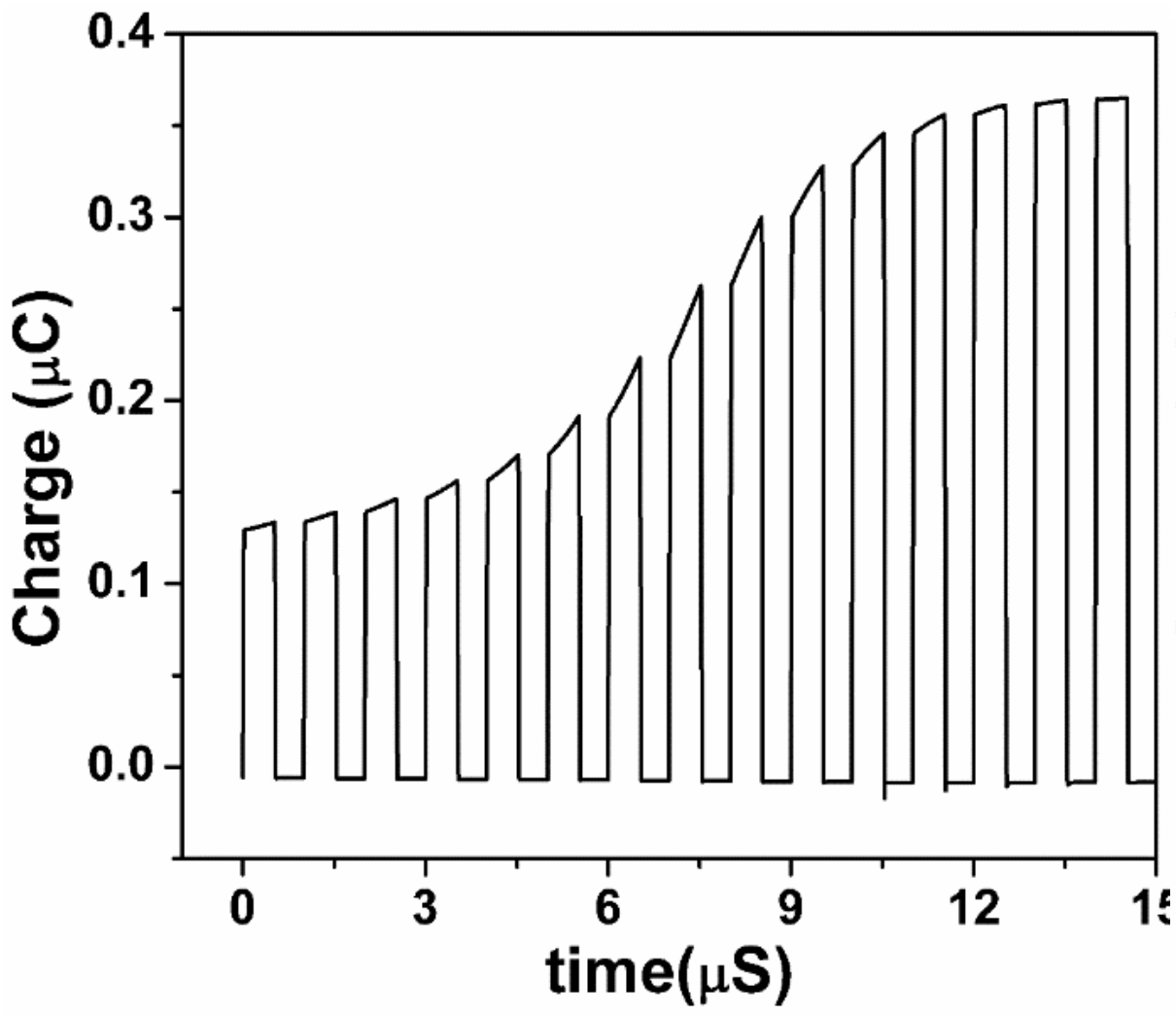

(a)

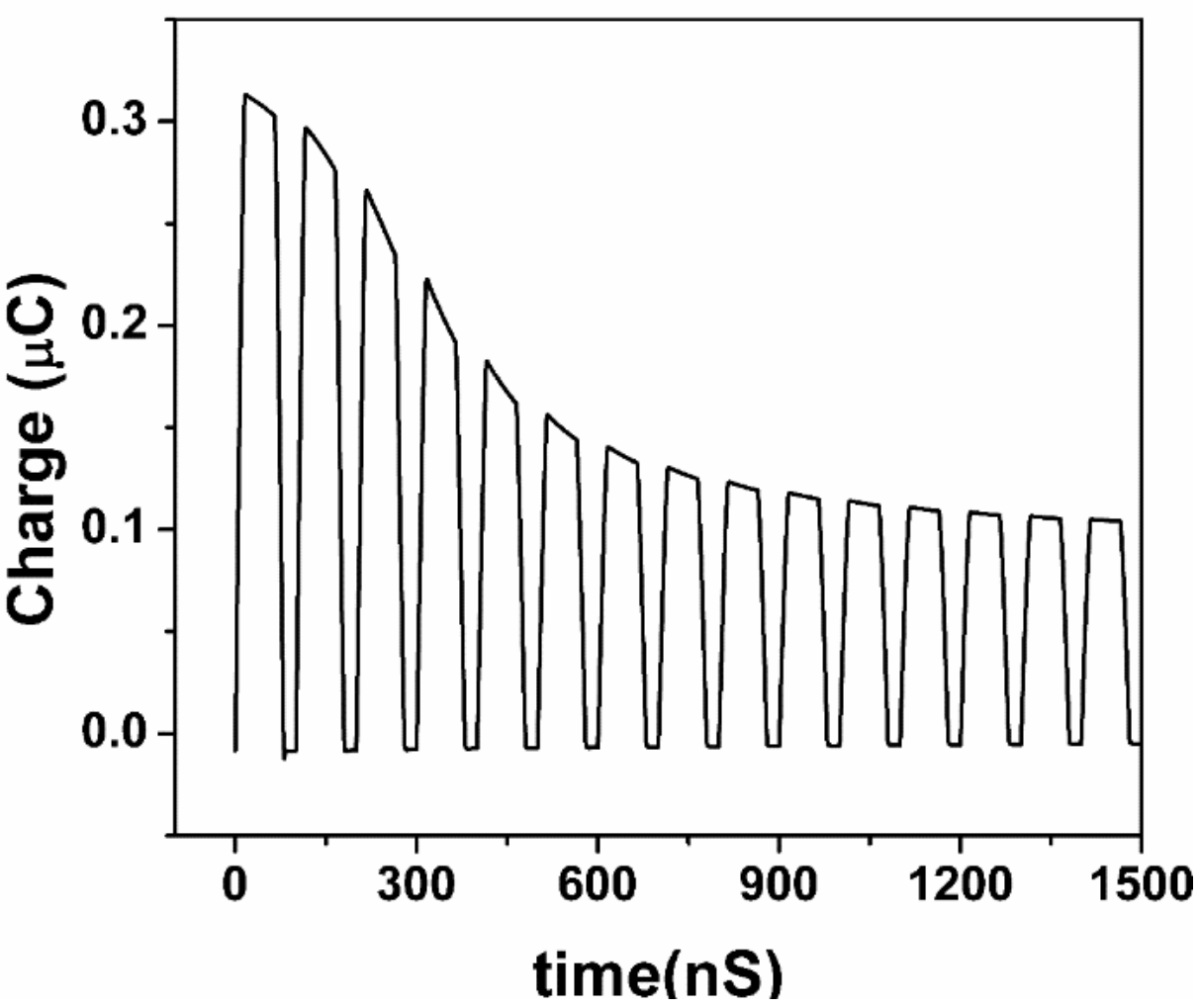

(b)

**Fig. 8.** Charge versus time plot for 350mv voltage pulse input for memcapacitor for $C_m$ $=2n, V_B=500m, t_p=1usec, t_w=500n$ sec,$C_2=500p$ at 500 KHz for (a) incremental topology, (b) decremental topology

## *5.3. Effect of variation of bias voltages on the pinched hysteresis loop*

It is seen in (15) and (16) that memcapacitor of emulators depends on transconductance $G_{m1,2}$ which is electronically controllable by external bias voltage $V_{B1,3}$ as per (2). Fig. 9(a) shows the simulation results for variation of bias voltage $V_{B1}$, $V_{B3}$ for signal frequency of 1 KHz, capacitor value of $C_1$=240nF, $C_2$=500nF , $A_m$=350 mV at different values of bias voltages (0.4V, 0.6V and 0.8V) . It is observed in Fig. 9(a-b) that the pinched hysteresis loop of q-v curves increases with the increase of $V_{B1,3}$ (or $G_{m1,3}$) as expected. It implies that the memcapacitance can be controlled by $V_{B1,3}$. It can further be noted that as $V_{B2}$ increases the pinched hysteresis loop of q-v rotates anticlockwise.

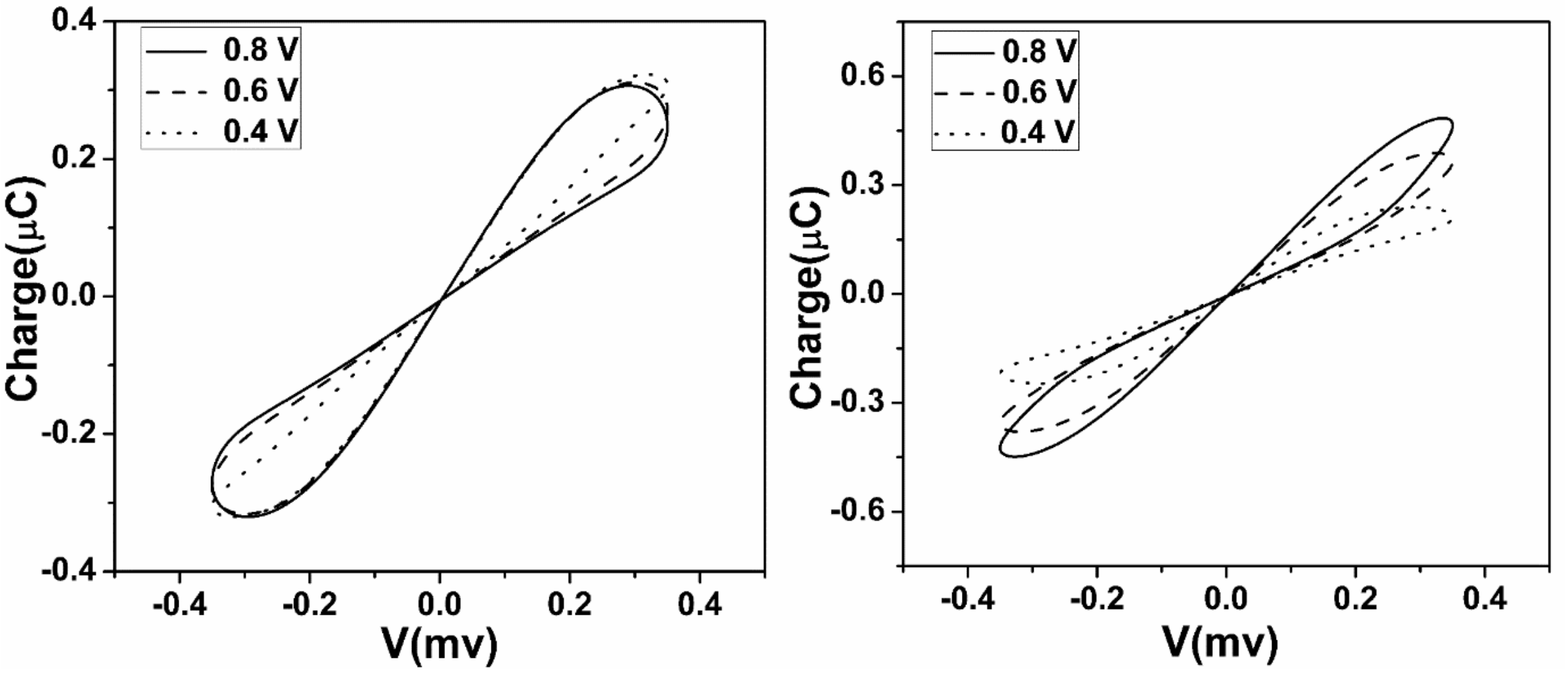

(a) (b)

**Fig. 9.** q-v Characteristic plot for (a) memcapacitor at 1 KHz for $V_{B1}$= 0.4, 0.6 and 0.8 V (b) memcapacitor at 1KHz for $V_{B1}$= 0.4, 0.6 and 0.8 V

## *5.5. Effect of capacitance and frequency variation on the pinched hysteresis loop*

The effect on q-v characteristics for variation of applied signal frequency for a fixed capacitance is shown respectively in Fig.10(a). Similarly, Fig. 10(b) show the effect of variation of capacitance for a fixed operating frequency.

Fig 10(a) shows the effect of varying frequency at a fixed capacitance of $C_1$=240nf, Am= 150 mV and $V_{B1}$ = 620 mV. It shows that as frequency increases from 1KHz – 1.5KHz for a fixed capacitance value of 240 nf, hysteresis loop becomes more and more linear which satisfies (15,16) suggesting that the time-varying nature of the loop decreases and ultimately vanishes at a certain frequency, satisfying the fingerprint of memcapacitor.

Fig 10(b) shows the effect of varying capacitance with a fixed frequency of 1KHz which are in conformity to (15) and (16). As the capacitance increases hysteresis loop shrinks and ultimately becomes less linear satisfying the fingerprints of memcapacitor.

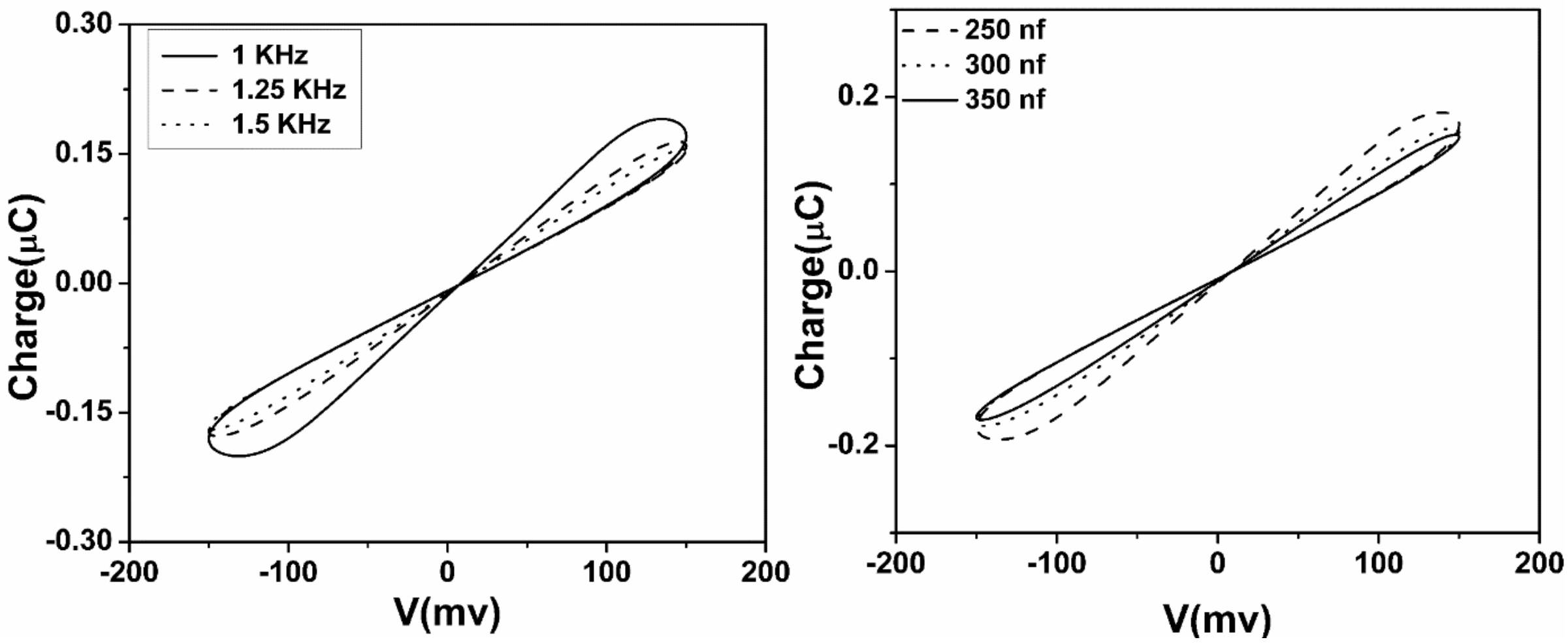

(a) (b)

**Fig.10.** The Q-V Characteristic plot for incremental configuration for (a) memcapacitance at a fixed frequency of 1KHz for variable capacitance from 250pf-350nf (b) memcapacitance at a fixed capacitance of 240nf for variable frequency from 1KHz-1.5KHz

## *5.6. Parallel combinations of memcapacitor*

Fig. 11 shows the memcapacitor in parallel combination. Fig 12(a) shows the equivalent and individual charge on memcapacitance $C_{M1}$, $C_{M2}$. As in parallel combination equivalent charge, $Q_{equ}$ becomes $Q_{M1}+Q_{M2}$ but corresponding voltages remains the same. Fig 12(b-c) shows the individual and combination hysteresis loop for memcapacitors in parallel. It can be noted that $C_{equ}$ hysteresis loop is approximately twice that of the individual which is similar to the addition of capacitors in parallel.

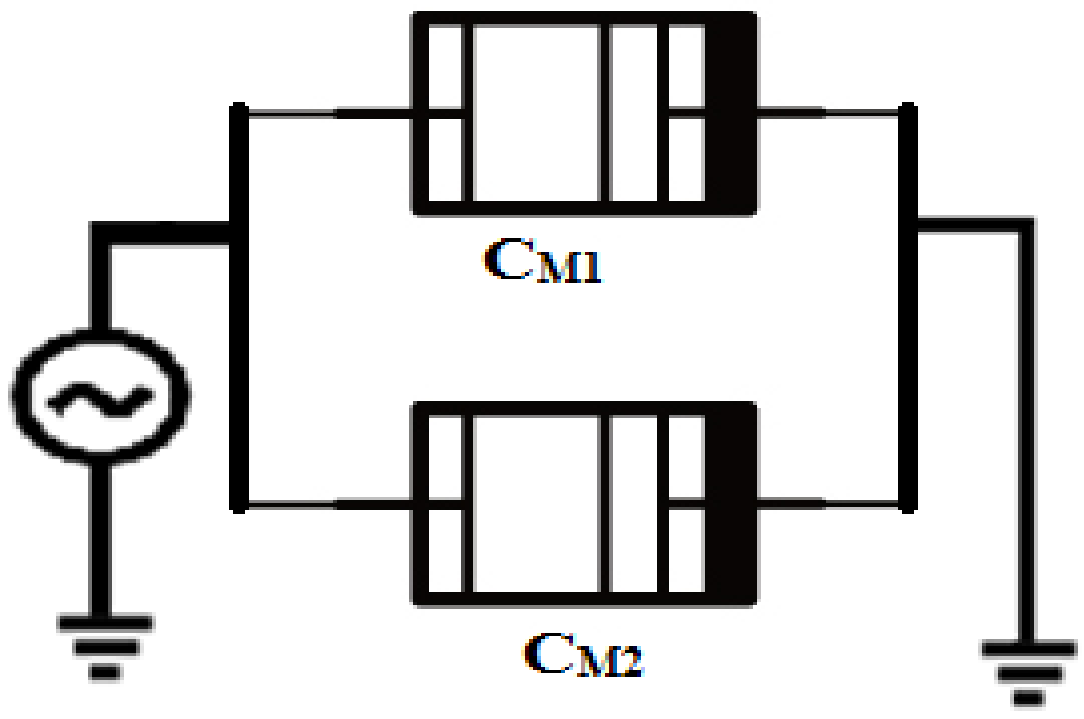

**Fig.11**. Memcapacitor $C_{M1}$, $C_{M2}$ in parallel combination

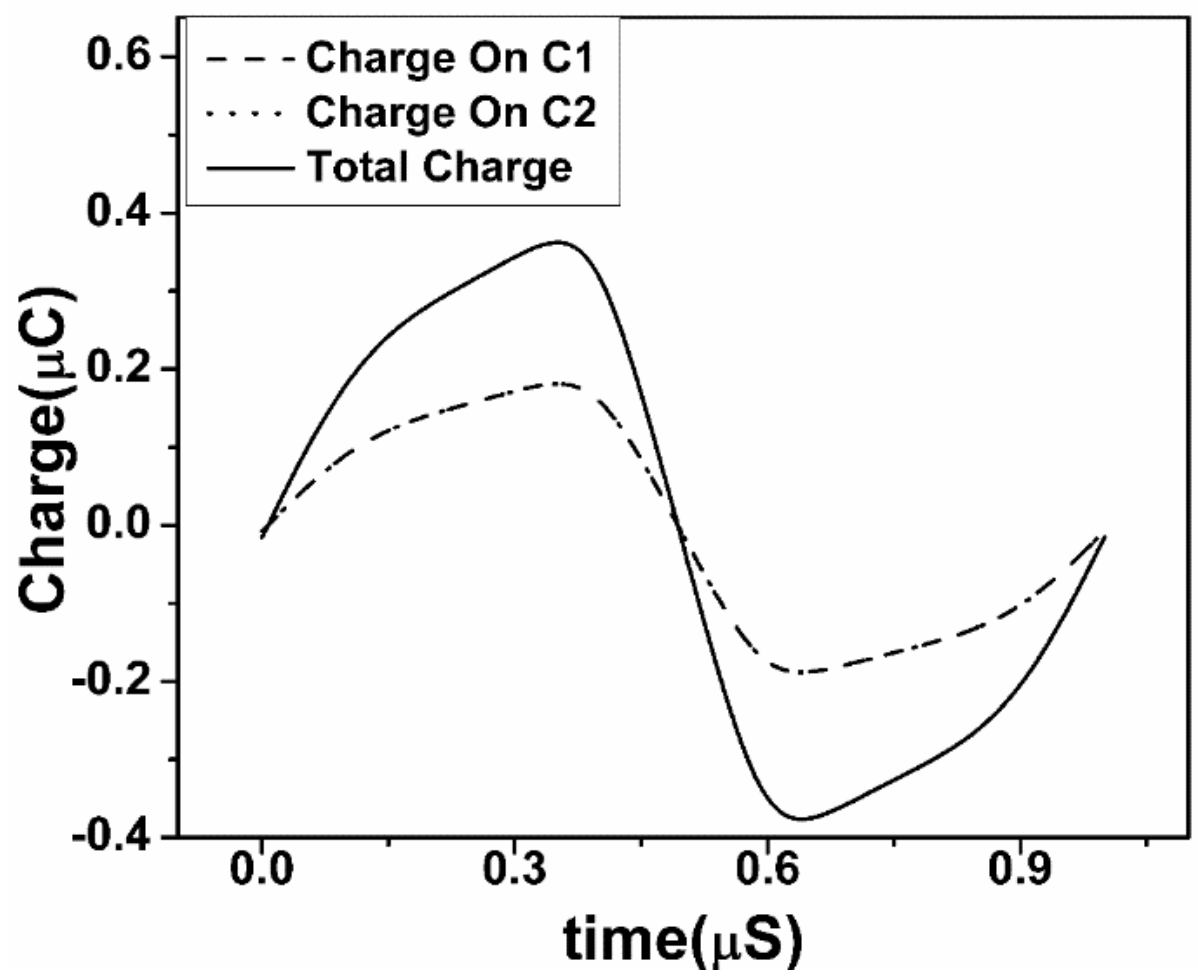

(a)

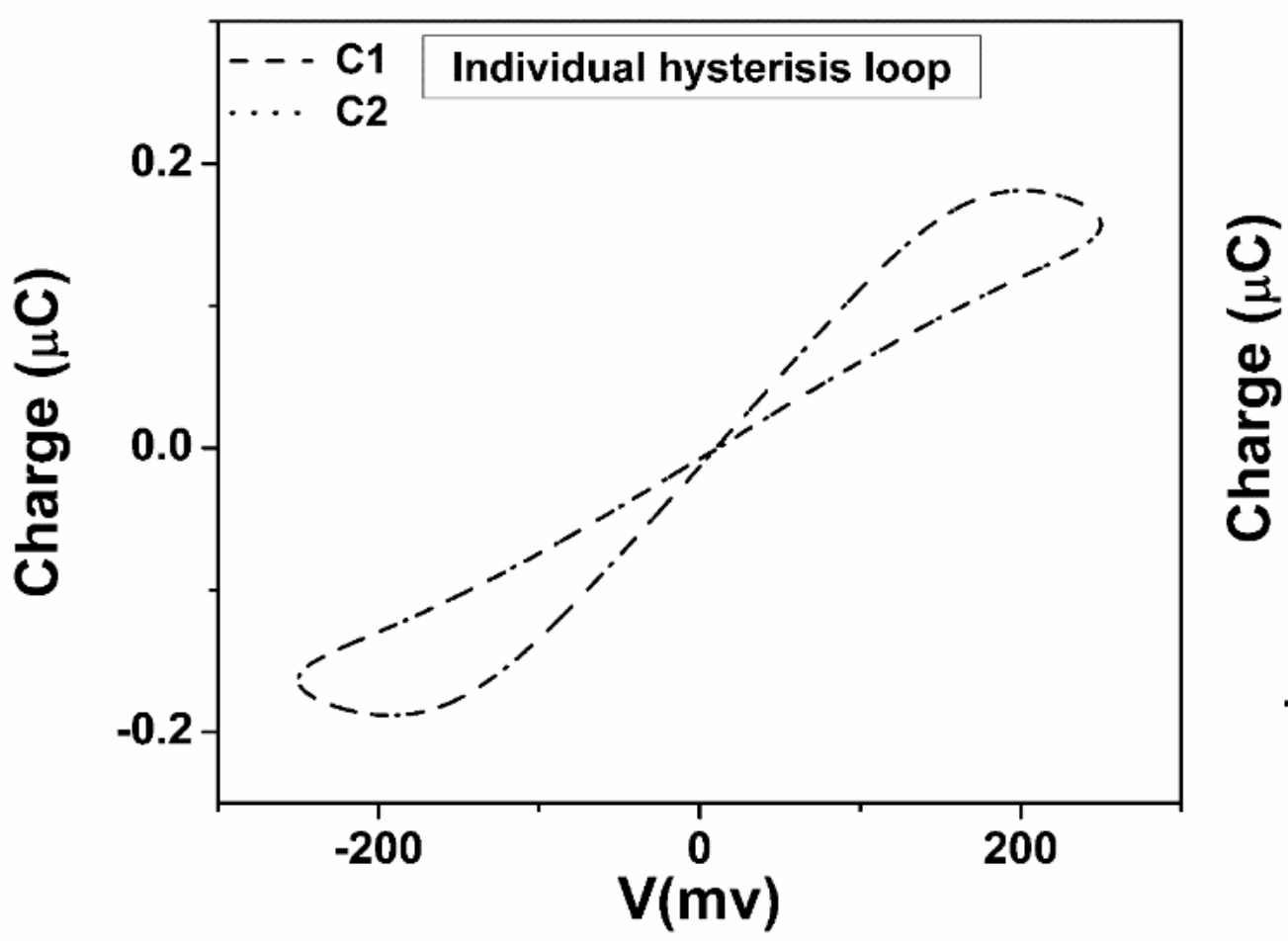

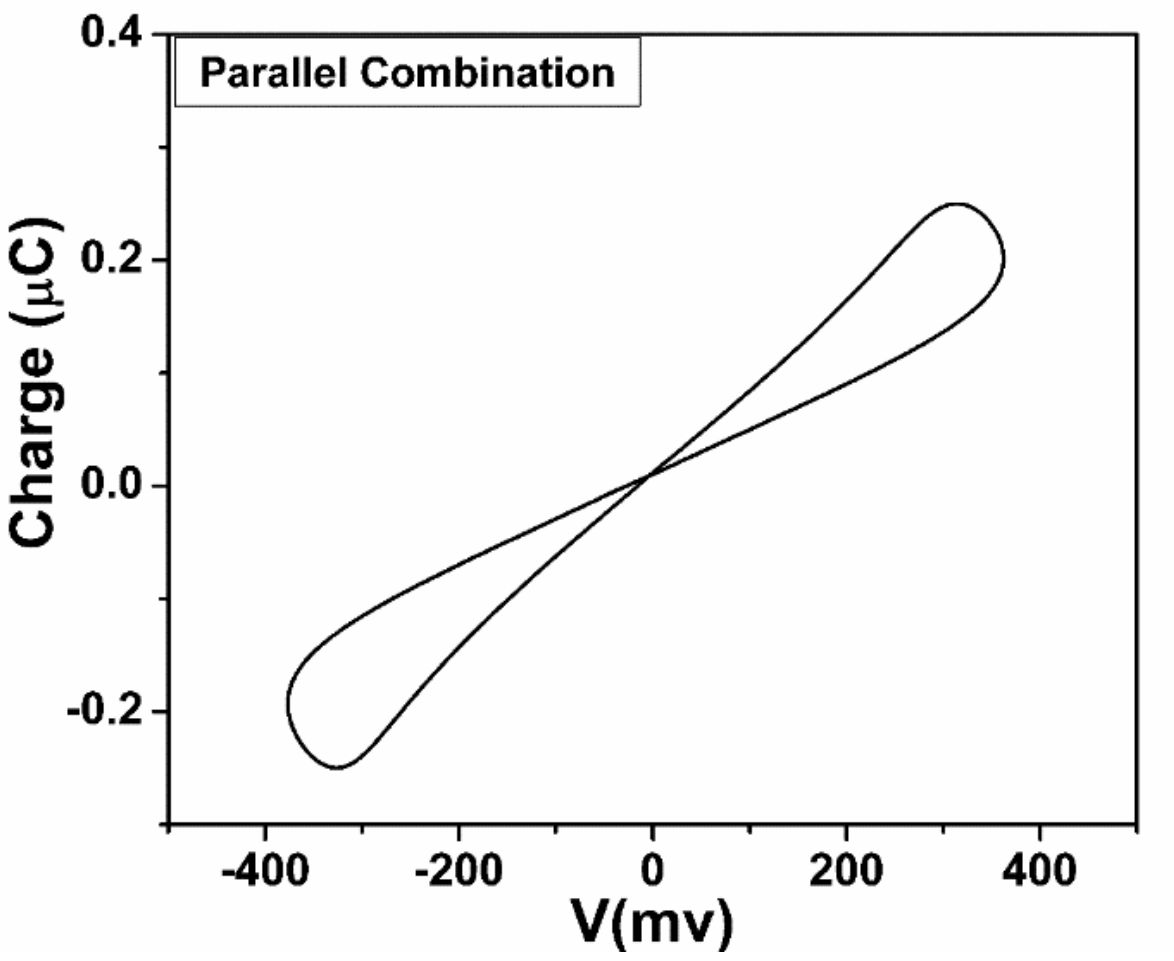

(b) (c)

**Fig. 12**. (a) charges vs time plot for individual and combinational parallel memcapacitance (Fig. 11) (b) q-v plot for $C_{M1}$, $C_{M2}$ memcapacitor for f=1MHz, C=120pf, Vs=250mv, Vb=400mv, (c)q-v plot for parallel memcapacitor for f=1MHz, C=120pf, Vs=250mv,Vb=400mv

# 6. Layout, post-layout simulations, and comparison

Layouts are obtained and post-layout simulations are carried out to check the effect of parasitic on the hysteresis and charge v/s voltage plots.

## *6.1. Layout*

Layouts of memcapacitor have been obtained for post-layout comparison. The view of the layout of the block of memcapacitor emulator is shown in Fig. 13.

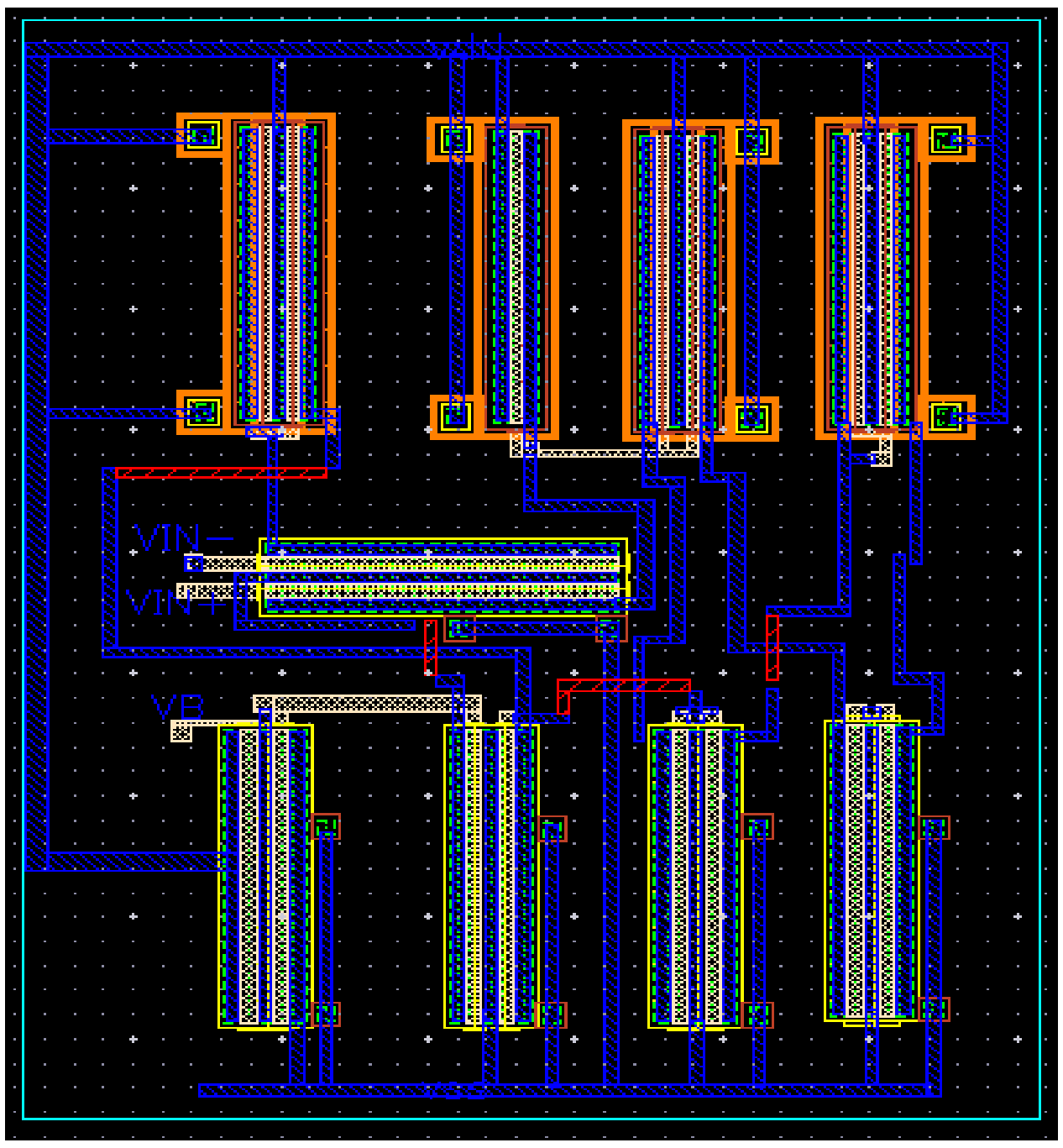

**Fig. 13.** Layout of analog block(OTA) of memcapacitor emulator Fig. 4

## *6.2. Pre and post-layout comparison for proposed memcapacitor emulator*

Fig 14(a-c) shows the Q-V characteristic plot for memcapacitance at 8 MHz, 5MHz, and 1MHz respectively. Simulations are carried out for a fixed FC=2×10-3 FaradHz and a bias voltage of 640mV.

It is observed that the pre and post layout results overlap with each other, which reveals that the effect of parasitic on the proposed circuit built using OTA and second generation Current Conveyor is negligible.

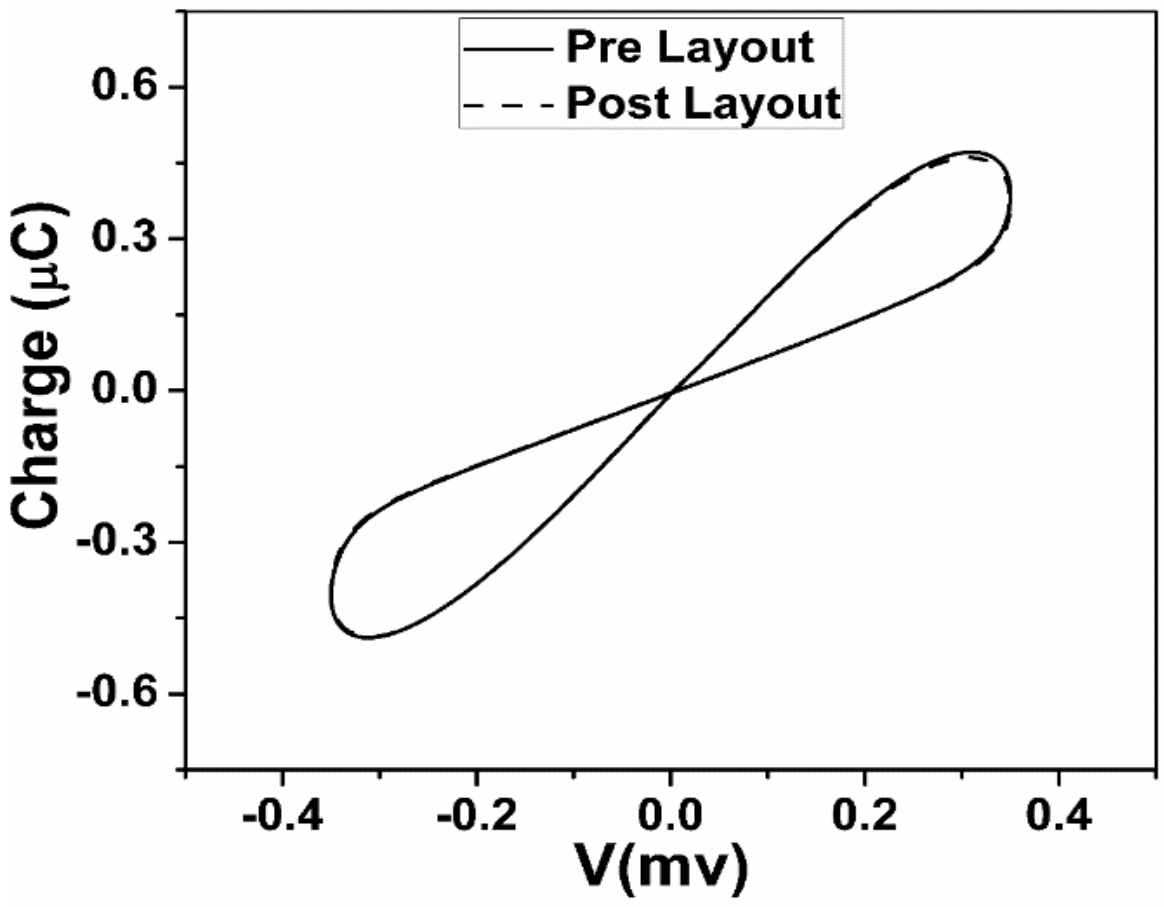

(a)

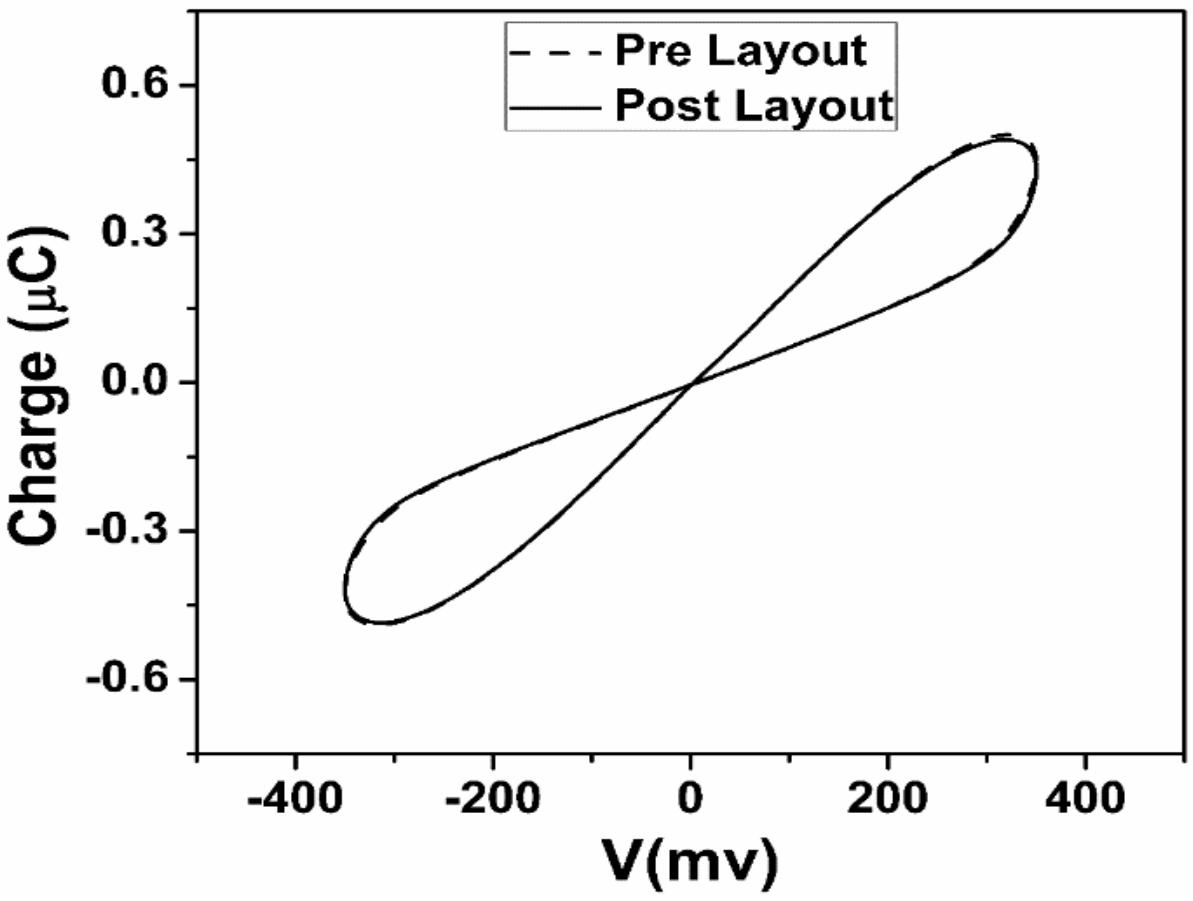

(b)

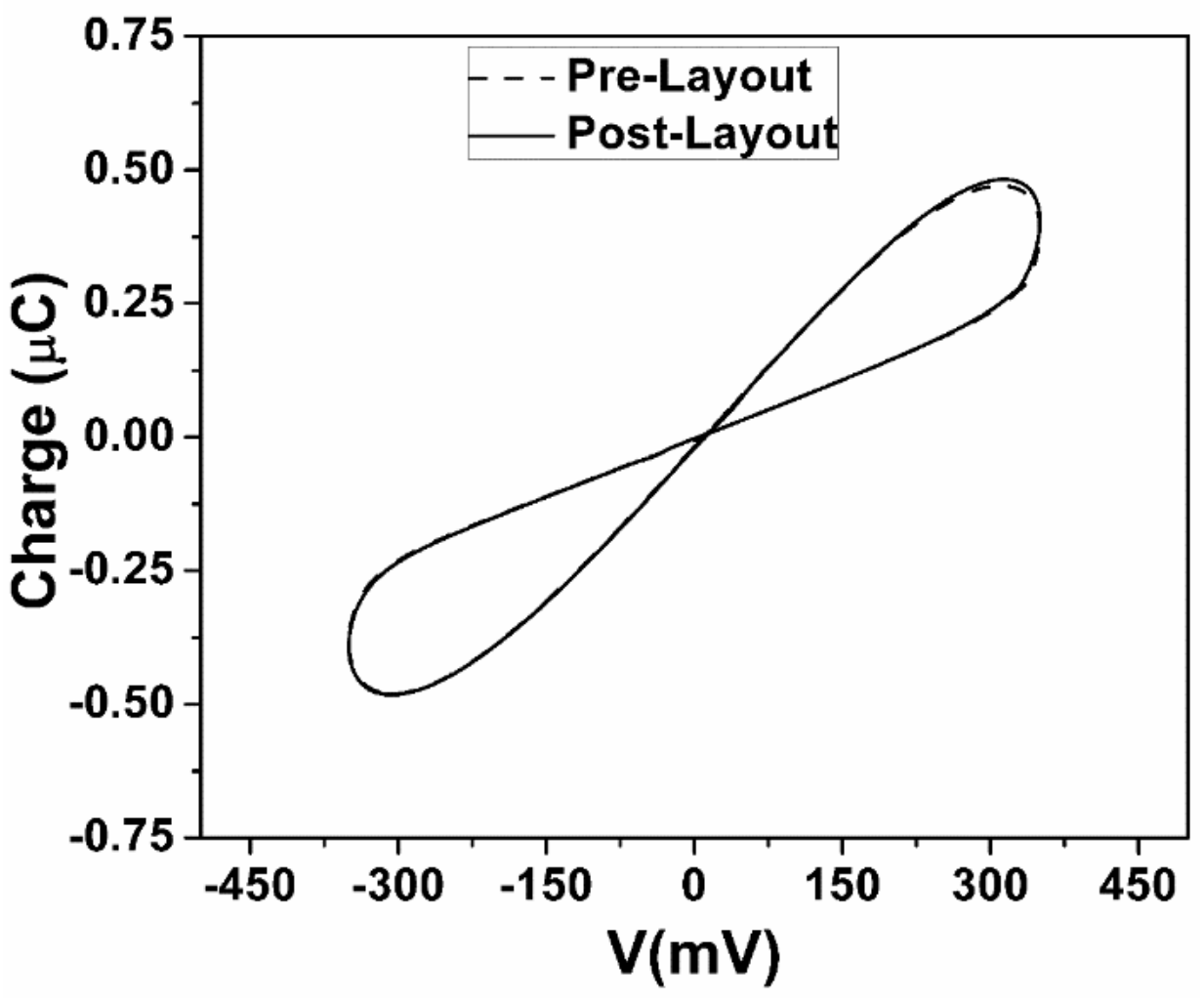

(c)

**Fig. 14.** Pre and Post layout comparison of Q-V characteristic plot (a) at 8 MHz, (b) at 5 MHz, (c) at 1 MHZ , operating frequency for $V_{B1,3}$= 0.6 V, Vin = 350 mV, and constant $C_1f$ =2×$10^{-3}$ FHz,

## *6.3. Monte Carlo Analysis*

Post layout Monte Carlo (MC) simulation for process mismatch at a frequency of 1KHz for 200 simulation runs is performed for decremental topology and the effect on hysteris loop is shown in Fig. 15. Gaussian random variation of standard device parameters is mentioned in Table 4.

**Table 4. Deviation values for process and mismatch**

| Parameters | Process Deviation | Mismatch Deviation |
|---|---|---|
| **tox** | 0.2e-9 | 0.02e-9 |
| **Vth** | 0.04 | 0.004 |
| **L** | 2e-9 | 0.2e-9 |
| **W** | 2e-9 | 0.2e-9 |
| **Cjn** | 0.00015 | 0.000015 |
| **Cjswn** | 0.3e-10 | 0.03e-10 |
| **Cjswgn** | 0.5e-10 | 0.05e-10 |
| **Cgon** | 0.6e-10 | 0.06e-10 |
| **hdifn** | 2e-8 | 0.2e-8 |

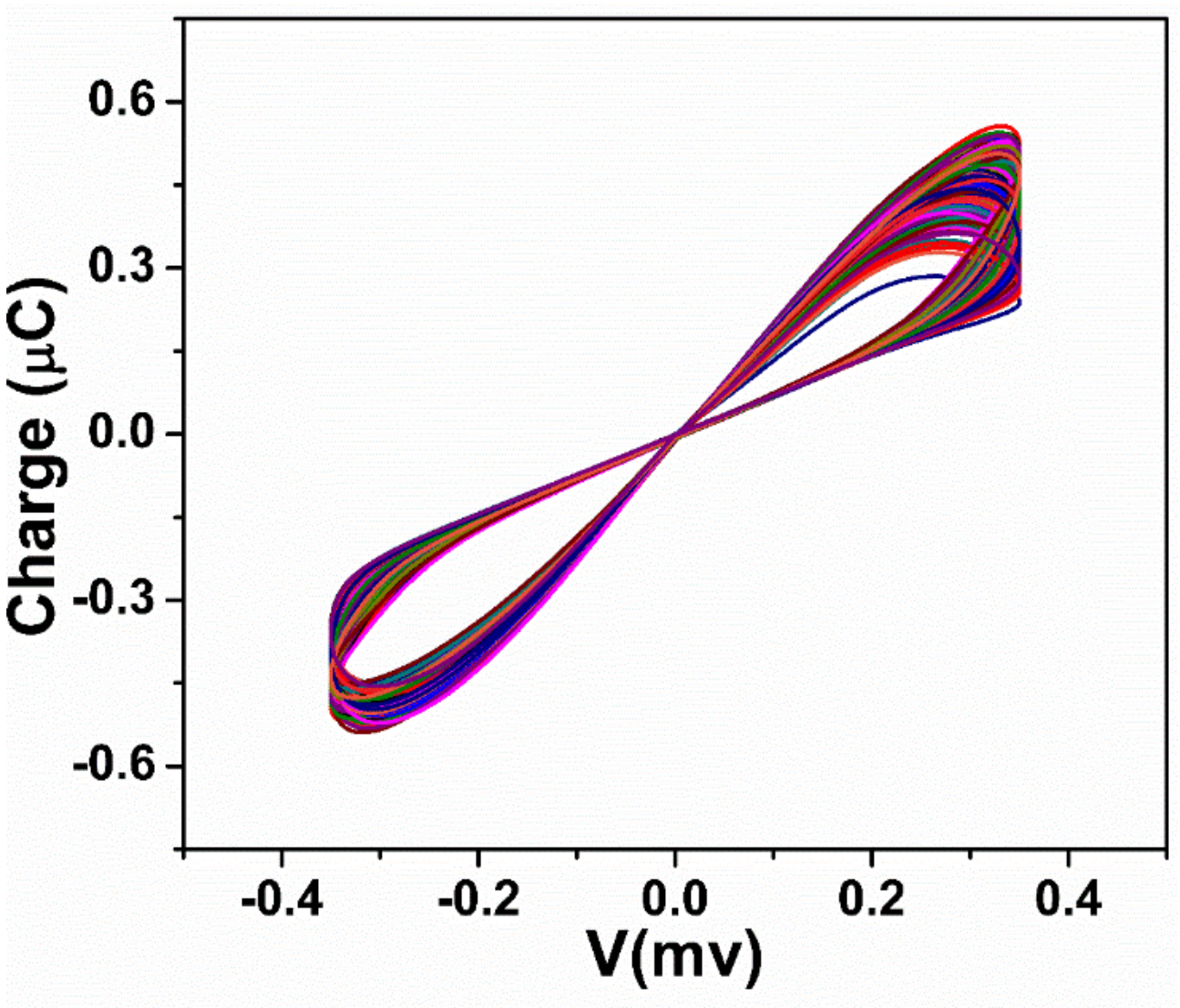

**Fig. 15.** Hysteresis loop of MC result for 200 simulation runs (a) for 1KHz sinusoidal input at Cap $C_1$ = 240nf, $V_{B1,3}$ =600mV

Fig.16 shows the histogram plot for distribution of samples for memcapacitance dependent parameters such as threshold voltage(Vth) and betaeff (K). From histogram graph, the standard deviation for Vth and K are found to be 32.4 mv and 416 u respectively for both topologies.

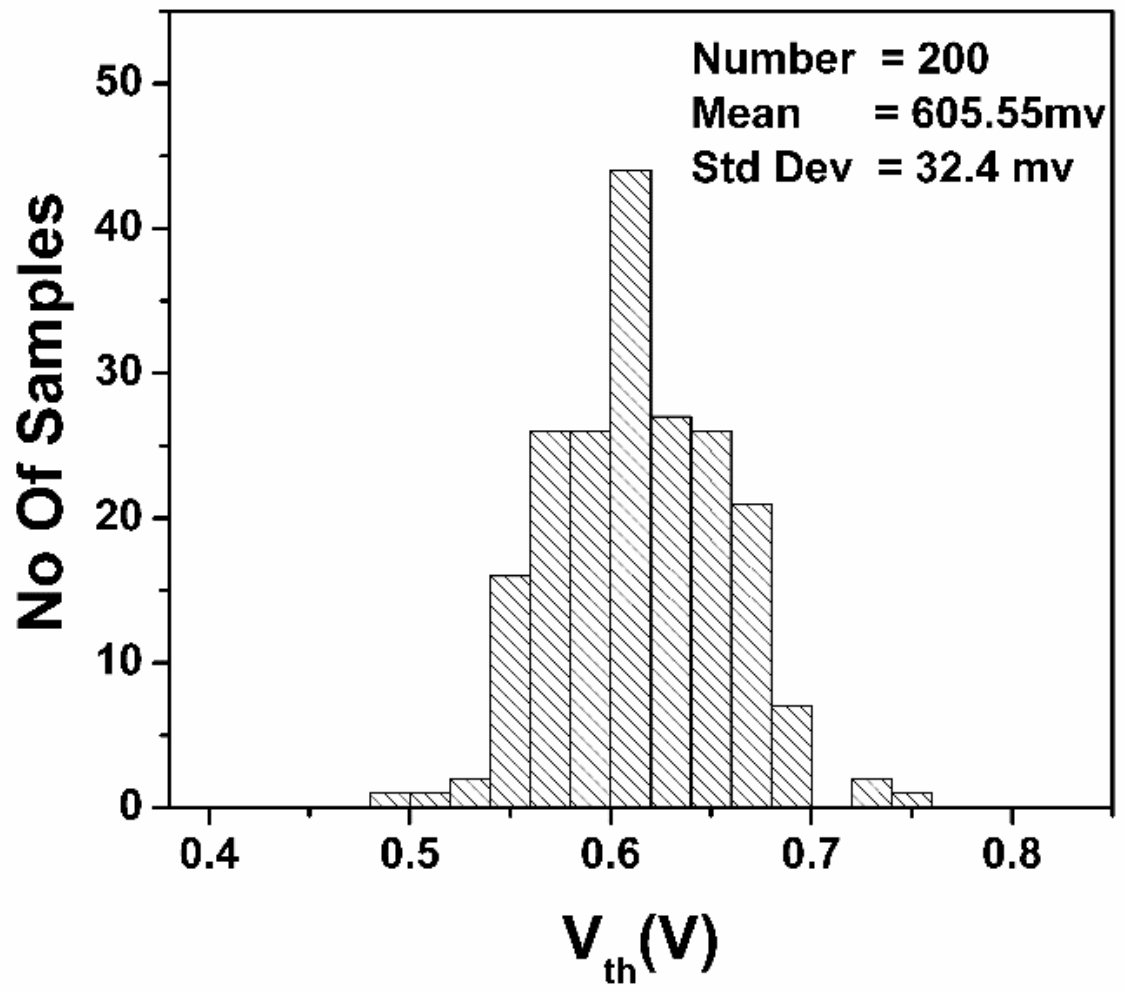

(a)

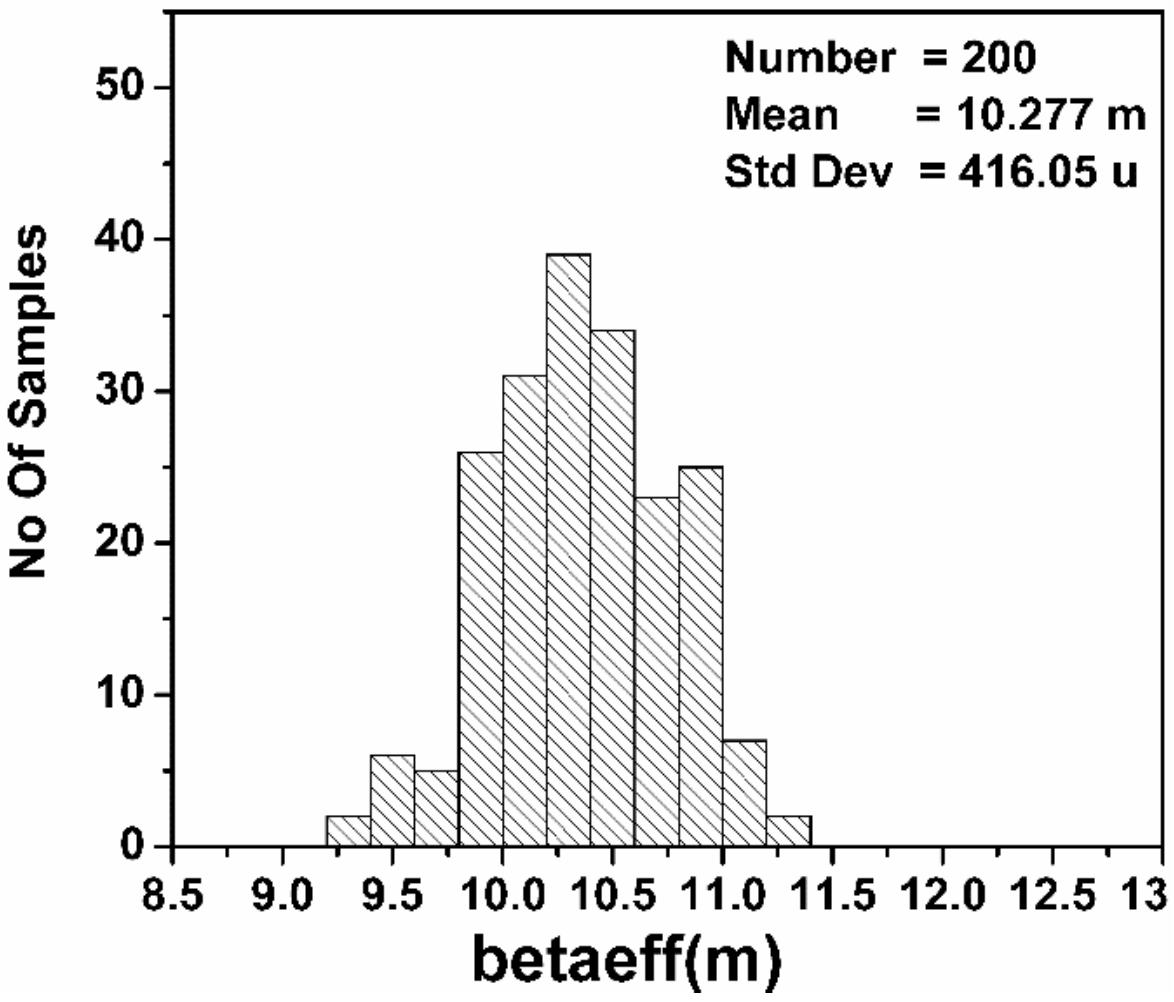

(c)

**Fig. 16.** MC result for 200 simulation runs for (a) threshold voltage (b) betaeffect

On analyzing Fig.16, it can be seen that Hysteresis loop shows a bunch of expected loops as memcapacitance depends on capacitance, parasitic and other device parameters. However, the loops remain pinched at the origin and the memristive nature is perfectly conserved. It reveals that the proposed circuit has reasonable sensitivity performance with a variation of values of process, mismatch, threshold voltage and other device parameters.

## 7. Non-Ideal Analysis

### 7.1 *Non-ideality effect due to OTA transconductance gain*

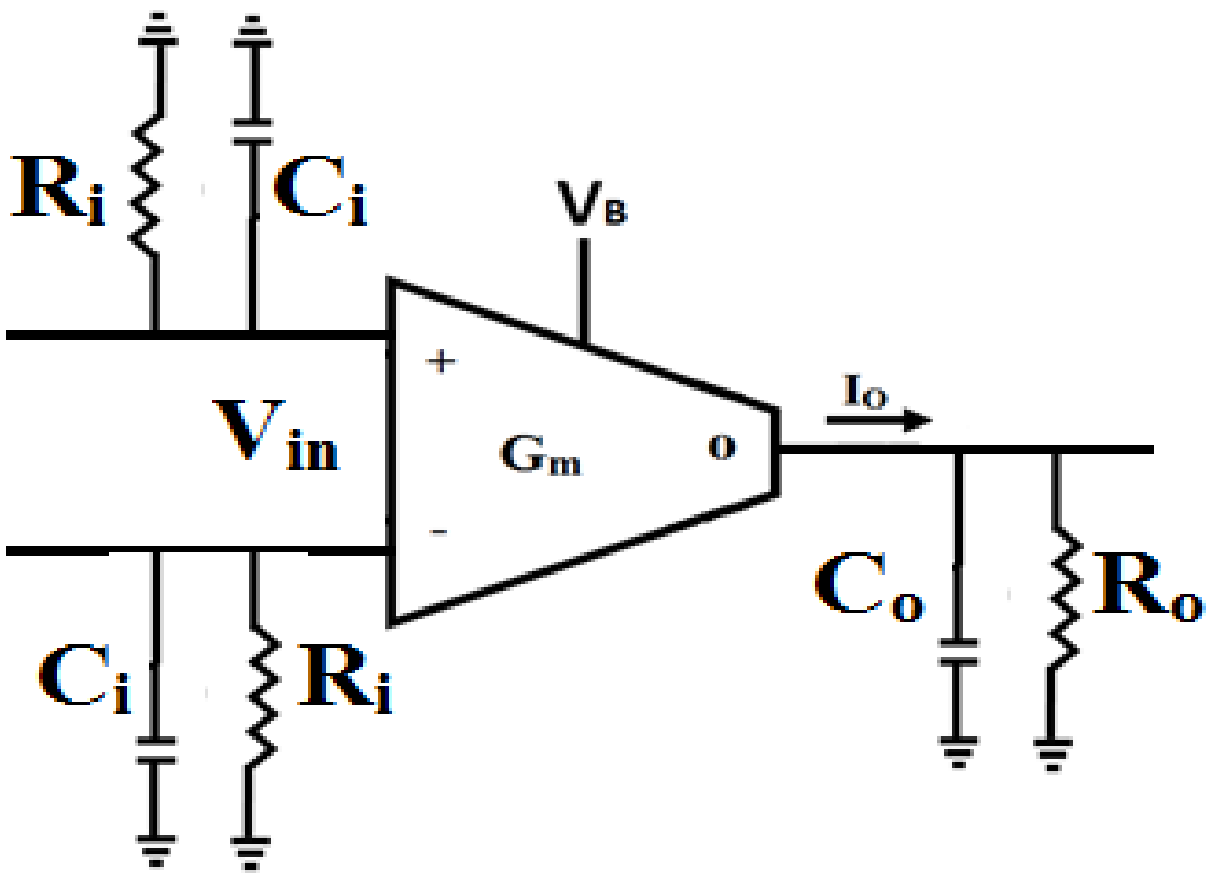

**Fig. 17.** Non-ideal model of OTA [33, 34]

Fig. 17 shows the non-ideal model of OTA where ($R_i$, $C_i$) and ($R_0$, $C_0$) are input and output parasitic capacitances and resistances respectively. The capacitances across the input ports are assumed to be equal. The transconductance of non-ideal OTA can be characterized by a single pole roll-off model [33-34] giving frequency dependent transconductance at low frequency as,

$$G_m(s) = G_{m0}\frac{\omega_a}{s+\omega_a} \tag{19}$$

where $G_{m0}$ is the low-frequency value or open loop transconductance and $\omega_a$ is the frequency in radians per second for which the transconductance has decreased to 0.707 of its low-frequency value or simply called the first corner frequency. At very high frequency the phase shift of OTA become a monotonically decreasing function of frequency which can be further approximated as [61]

$$G_m(s) = G_{m0}\frac{\omega_a}{s+\omega_a}e^{-s\tau} \;\; , \;\; s = j\omega \tag{20}$$

where $\tau$ denotes excess phase shift whose approx. practical value being 1.25ns [33].

Equation (19) and (20) can be written as

$$G_m(s) = \gamma G_{m0} \text{ ,where } \begin{pmatrix} \gamma = \dfrac{\omega_a}{s+\omega_a}, & \textit{for} \text{ low and mid frequency} \\ \gamma = \dfrac{\omega_a}{s+\omega_a} e^{-s\tau}, & \textit{for} \text{ very high frequency} \end{pmatrix} \tag{21}$$

Taking into account the non-ideal effects in CMOS implementation of OTA the modified port relationship can then be written as

$$I_0 \pm = \pm \gamma G_{m0}(V_{IN+} - V_{IN-}) = \pm G_m V_{in} = \pm \gamma G_{m0} V_{in} \tag{22}$$

Where trans-conductance gain coefficient from the input terminal to output terminal of OTA is denoted by $\gamma$, which is ideally taken to be unity. Re-analyzing the memcapacitance emulator circuits of Fig. 4 and assuming all the OTA have $\gamma$ as the transconductance gain coefficient, the memcapacitance expression (15) and (16) are modified as

$$\frac{I_{in}}{\mathrm{V}_{in}} = S\,\mathrm{C}_3 \left[ 1 + \frac{G_{\mathrm{m}3}}{\dfrac{k}{\sqrt{2}} \left( V_{ss} + 2V_{th} \mp \dfrac{\gamma G_{m1} G_{\mathrm{m}3} \phi_{in}}{G_{\mathrm{m}2} C_1} \right)} \right] \tag{23}$$

Equation (23) clearly indicates that the trans-conductance gain coefficient developed due to non-ideal effects in OTA changes memcapacitance value, however, the effect of nonideality may be neglected provided the operating frequency is much lower than the 3dB frequency $\omega_a$ of OTA.

## *7.2 Nonideality due to device parasitic*

Fig. 18. shows the non-ideal model with device parasitic of proposed memcapacitor emulators. Resistance($R_3$) and capacitance($C_3$) at node A in Fig.18 represent the equivalent input parasitics resistances and capacitances at the nodes. At the node B, Resistance $R_{02}$ is the parallel

combination of input parasitic resistances of two OTAs and output parasitic resistance of OTA 2. Similarly, capacitance $C_{02}$ is the parallel combination of input parasitic capacitance of two OTAs and the output parasitic capacitance of OTA 2. Rs being the source resistance. $R_{01}$ and $C_{equ}$ at node B in Fig.18 represent the output parasitic resistance and the parallel combination of output capacitance with capacitance($C_1$).

### *7.2.1. Non-ideal model of memcapacitor emulator due to device parasitics*

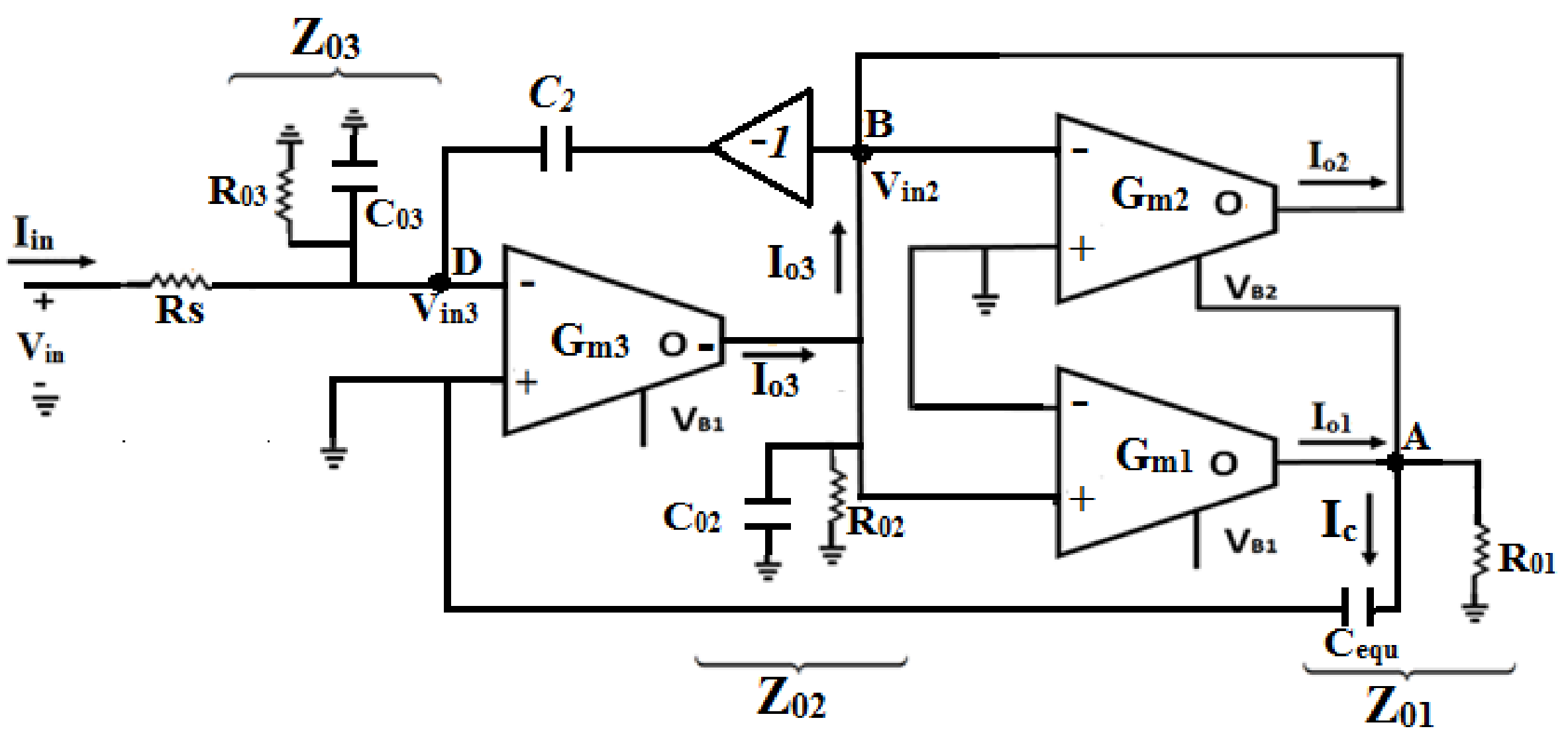

**Fig. 18.** Non-ideal model of proposed memcapacitor emulator

Let the equivalent impedance at node 'A', node 'B' and node 'D' be $Z_{01}$ , $Z_{02}$ and $Z_{03}$ respectively where

$$Z_{01} = \left(R_{01} \,||\, Xc_{equ}\right) \text{ and } Z_{01,02} = \left(R_{02} \,||\, Xc_{02}\right) , Z_{03} = \left(R_{03} \,||\, Xc_{03}\right)$$

and $$Xc_{eq} = \frac{1}{j\omega(c_1 + c_o)}, Xc_{02} = \frac{1}{j\omega(c_i + c_i + c_o)}, Xc_{03} = \frac{1}{j\omega c_i}, R_{02} = R_i \,||\, R_i \,||\, R_o \,.$$

On applying KCL at node D and node B respectively we get

$$V_{in}(t) = V_{in3}(t)\beta_1 \tag{24}$$

$$V_{in} = -\frac{Gm_2\beta_1}{Gm_3}V_{in2} \tag{25}$$

where

$$\beta_1 = \left[ S_{C2} - S_{C2}\frac{G_{m3}}{G_{m2} + \frac{1}{Z_{02}}} + \frac{1}{Z_{03}} + \frac{1}{R_S} \right] R_s$$

Also $$V_{B2} = \frac{1}{C_{eq}}\int I_c dt = \frac{R_{01}G_{m1}}{Z_{01}C_{eq}}\int V_{in2} dt = \frac{R_{01}G_{m1}G_{m3}}{Z_{01}G_{m2}C_{equ}\beta_1}\phi_{in} \qquad (26)$$

On applying KCL at node D and using (24), (25), (26) we get

$$\frac{I_{in}}{V_{in}} = \frac{1}{\beta_1 Z_{03}} + \frac{S\,C_3}{\beta_1}\left[ 1 + \frac{G_{m3}}{\frac{k}{\sqrt{2}}\left( V_{ss} + 2V_{th} + \frac{R_{01}G_{m1}G_{m3}\phi_{in}}{Z_{01}G_{m2}C_{equ}\beta_1} \right)} \right] \qquad (27)$$

Similar analysis for decremental topology results in

$$\frac{I_{in}}{V_{in}} = \frac{1}{\beta_1 Z_{03}} + S\,C_3\left[ \frac{1}{\beta_1} + \frac{G_{m3}}{\frac{k}{\sqrt{2}}\left( \beta_1\left(V_{ss} + 2V_{th}\right) - \frac{R_{01}G_{m1}G_{m3}\phi_{in}}{Z_{01}G_{m2}C_{equ}} \right)} \right] \qquad (28)$$

Typical numerical values of CMOS OTA parasitic obtained from routine analysis and [35] can be assumed approximately as $R_{in} = \infty,\ R_o = 1M\Omega,\ C_i = 50\,fF,\ C_0 = 100\,fF$ which gives $\beta_1 \approx 1$ at high frequency and low frequency for $G_{m3} = G_{m2}$. Also, $Z_{01} \simeq R_{01}$ at low and high frequency and $Z_{02}, Z_{03}$ being $\infty$ and high at low and high frequency respectively which on further substitution in (27) and (28) gives

$$\frac{I_{in}}{V_{in}} = \frac{S\,C_3}{\beta_1}\left[ 1 + \frac{G_{m3}}{\frac{k}{\sqrt{2}}\left( V_{ss} + 2V_{th} \pm \frac{R_{01}G_{m1}G_{m3}\phi_{in}}{Z_{01}G_{m2}C_{equ}\beta_1} \right)} \right] . \qquad (29)$$

This is similar to (14) and (16). Hence it can be concluded that the effect of parasitics on the circuit is negligible for both high and low frequency of operation.

## 8. Application of memcapacitor as an amplitude modulator (AM), Filter and Oscillator

### *8.1 Application as Amplitude Modulator*

As an application of the proposed memcapacitive device, an AM modulation scheme with memcapacitor is carried out. Schematic of various circuits along with implementation using OTA based memcapacitor emulator is shown in Fig. 19. In Fig. 19(a) a multifunction filter [36] using OTA is given which can implement both bandpass filter (BPF) and low pass filter (LPF) responses using the components for $Y_1$, $Y_2$, $Y_3$, $Y_4$ and $Y_5$ as given in Table 5. The memcapacitance of memcapacitor shown in Fig. 19(b) is controlled by the low-frequency message signal $V_m(t)$ due to which message gets imposed on the high-frequency carrier signal $V_c(t)$. The output is filtered out by the bandpass filter of Fig. 19(a) centered at the carrier frequency to obtain an Amplitude Modulated wave. The circuit of Fig. 19(c) is used to demodulate the AM signal to recover message signal.

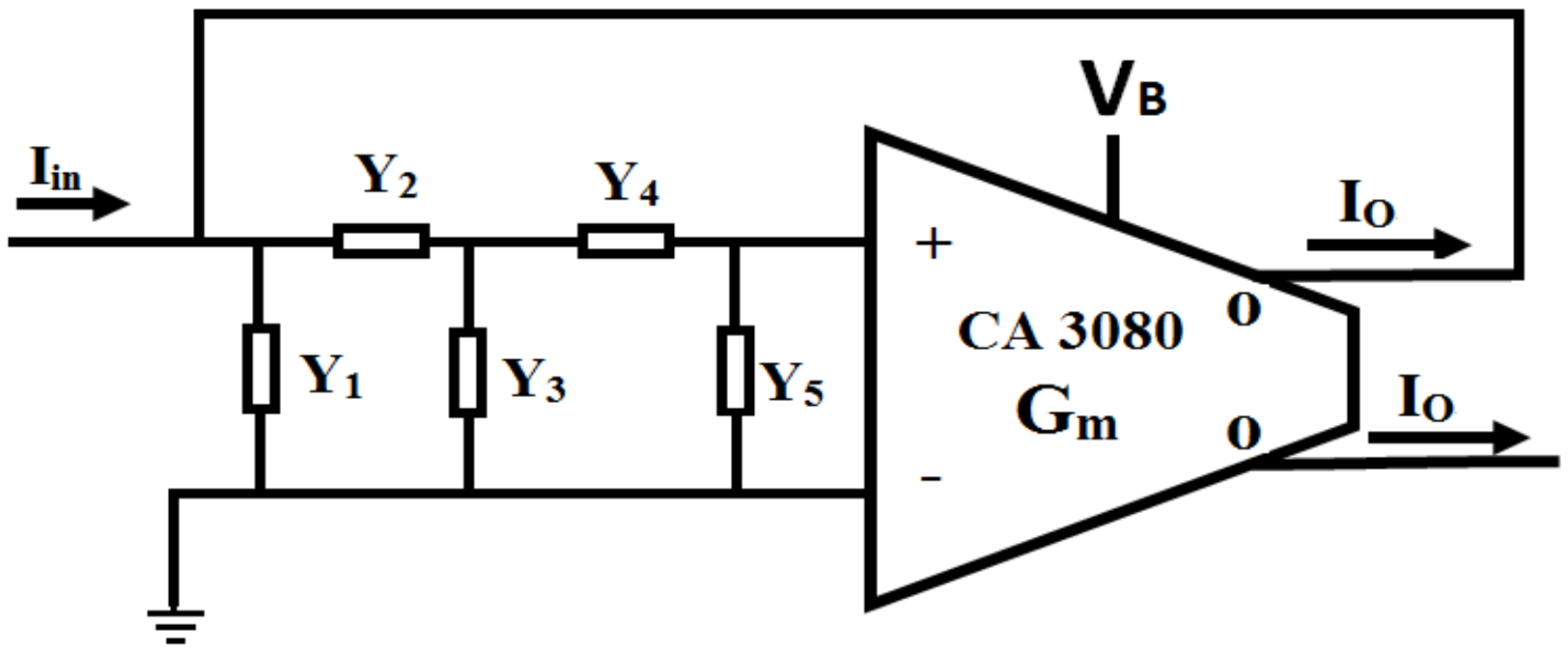

(a)

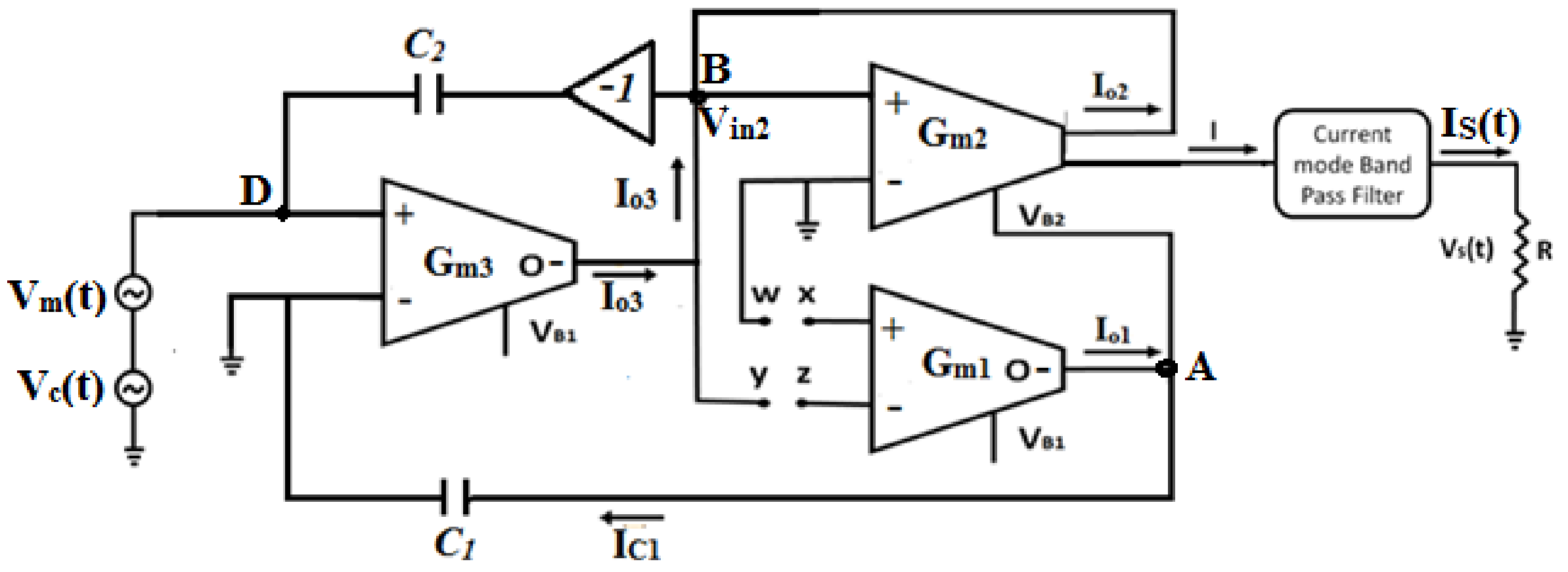

(b)

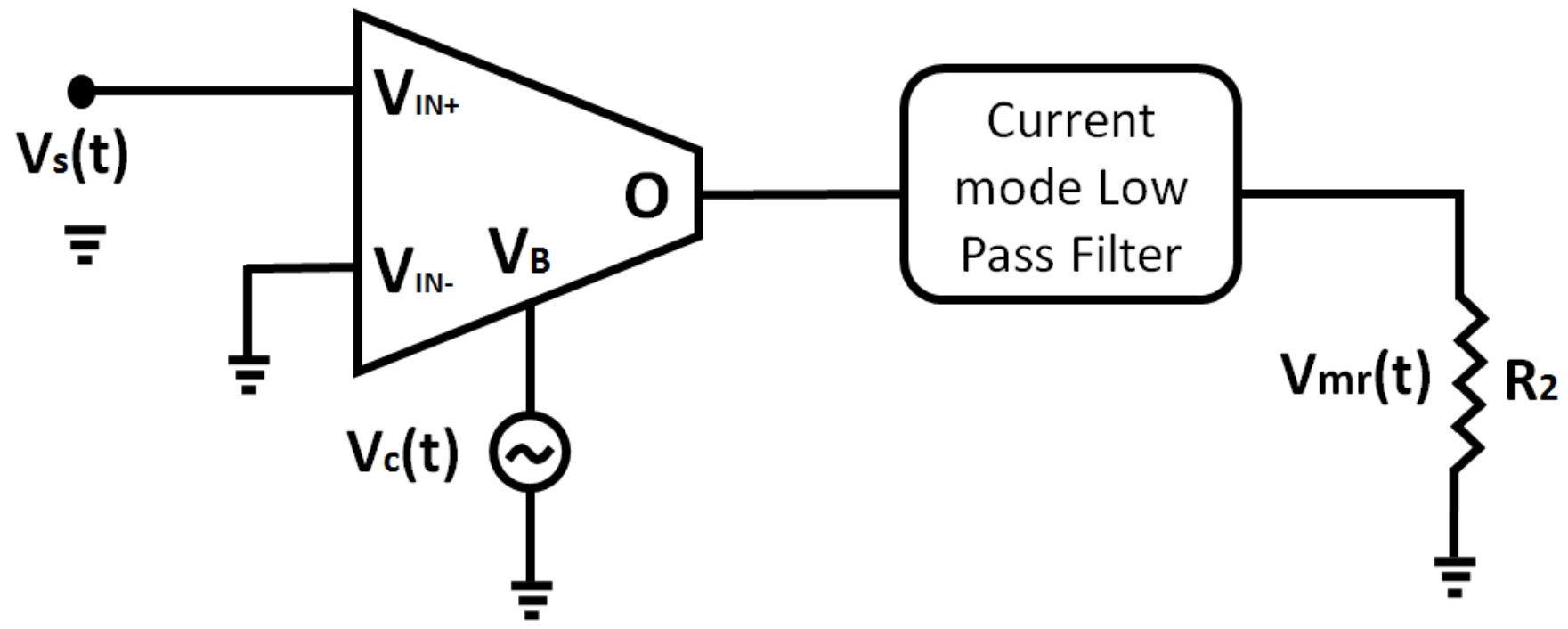

(c)

**Fig.19.** Block diagram of (a) current mode multifunction filter, (b) AM modulator circuit using memcapacitor emulator and filter, (c) coherent demodulator circuit using OTA

**Table 5.** Specification for multimode filter components

| | $Y_1$ | $Y_2$ | $Y_3$ | $Y_4$ | $Y_5$ |
|---|---|---|---|---|---|
| Low pass | C1 | R2 | C3 | $\infty$ | 0 |
| Band pass | R1 | R2 | C3 | C4 | R5 |

## *8.2 Analysis of amplitude modulator and demodulator*

The voltage message signal $V_m(t)$ and carrier signal $V_C(t)$ are taken respectively as

$$V_m(t) = A_m \cos(\omega_m t); \qquad V_C(t) = A_C \cos(\omega_c t) \tag{30}$$

The flux imposed to memcapacitor will be given by

$$\phi_{in}(t)=\int v_{in}(t)dt=\int[V_m(t)+V_C(t)]dt=\frac{A_m}{\omega_m}\sin(\omega_m t)+\frac{A_c}{\omega_c}\sin(\omega_c t) \tag{31}$$

For high value of carrier frequency $\omega_c$, the flux (30 ) reduces as

$$\phi_{in}(t)=\frac{A_m}{\omega_m}\sin(\omega_m t) \tag{32}$$

When the memcapacitor is operated in incremental topology, on applying KCL at node B and D and using (30), (31) and (32) results in

$$\frac{I_{02-}}{\mathrm{V}_{in}}=\frac{G_{m3}K}{G_{m2}\sqrt{2}}\left(-\frac{G_{m1}}{C_1}\int V_{in}(t)\,\mathrm{dt}-\mathrm{V_{SS}}-2\,\mathrm{V_{Th}}\right)$$

or
$$I_{02-}=\frac{G_{m3}K}{G_{m2}\sqrt{2}}\left(\frac{G_{m1}G_{m3}}{C_1}\int V_{in}(t)\,\mathrm{dt}-\mathrm{V_{SS}}-2\,\mathrm{V_{Th}}\right)\left(A_c\cos(\omega_c t)+A_m\cos(\omega_m t)\right) \tag{33}$$

After bandpass filtering of current at the carrier frequency, the AM output current [$I_s(t)$] is obtained as

$$I=K_1\sin(\omega_c+\omega_m)+K_1^{'}\sin(\omega_c+\omega_m)+K_2\sin(\omega_c-\omega_m)+K_3\cos\omega_c t \tag{34}$$

where

$$K_3=-\frac{A_c G_{m1}G_{m3}}{\omega_c C_1}$$
$$K_2=\frac{-A_m G_{m3}KG_{m1}G_{m3}A_c}{2G_{m2}\sqrt{2}C_1\omega_c}$$
$$K_1=\frac{G_{m3}^2 KG_{m1}A_m A_c}{G_{m2}2\sqrt{2}C_1\omega_m}$$
$$K_1^{'}=\frac{A_m G_{m1}KG_{m3}^2}{2\sqrt{2}G_{m2}C_1\omega_c}$$

(34) contains upper sideband, lower sideband, and Carrier signal frequency. In order to recover message signal from the modulated signal, a coherent product demodulator is used.

Demodulator circuit is realized with OTA as multiplier cascaded with OTA based low pass filter as shown in Fig 19(c).

### *8.3 . Simulation of amplitude modulator and demodulator*

The parameters used for simulation of amplitude modulator (AM) and demodulator are given in Table 6. Fig.20(a) shows the modulated message signal $V_S(t)$ obtained at the output of current mode bandpass filter in Fig 19(b) and the message signal applied to AM. Fig.20(b) shows the spectrum of the modulated signal obtained by applying rectangular window function present in FFT mode. Fig.20(c) shows the recovered message signal obtained at the output of current mode low pass filter in Fig 19(c). It confirms that the scheme of AM proposed in this section using memcapacitor circuit can satisfactorily be utilized. It can further be shown that by varying amplitude of carrier and message signal one can obtain under, over and critical modulations.

Fig.20(d) and Fig.20(e) shows amplitude modulated waveform and its hysteresis loop for one-time period respectively. On analyzing Fig. 20(d) and Fig. 20(e) simultaneously one can observe that the amplitude of hysteresis loop formed between charge and voltage increases and decreases according to increase and decrease in amplitude of message signal. So, the hysteresis loop itself changes its shape, size, and amplitude according to message signal.

For the first half cycle of the message signal i.e. for time period A-B in Fig.20(d), the amplitude of hysteresis loop in Fig. 20(e) decreases from point X to Y in the positive half cycle and point P to Q in negative half cycle of the modulated wave simultaneously. This results in mapping of message amplitude on the carrier signal.

Similarly, for the second half cycle of the message signal i.e. for time period B-C in Fig.20(d) hysteresis loop of Fig. 20(e) increases from point Y to X for the positive half cycle and point Q to P for the negative half cycle of the modulated wave simultaneously leading to the mapping of message amplitude on the carrier signal. Thus, hysteresis loop which is nothing but memcapacitance slope is freezed as the peak amplitude variation of carrier i.e. it follows the amplitude of message signal. From Fig .20(e) we find that the hysteresis loop overlaps each other on each half cycle.

**Table 6. AM circuit simulation parameter**

| S.No. | Parameter | Value |
|---|---|---|
| 1 | Message signal amplitude Am | 120 mV |
| 2 | Message signal frequency fm | 60 KHz |
| 3 | Carrier signal amplitude Ac | 370 mV |
| 4 | Carrier signal frequency fc | 2 MHz |
| 5 | Band Pass filter center frequency | 2 MHz |
| 6 | Band Pass Resistance ($R_1$=$R_2$=$R_5$) | 200 Ω |
| 7 | Band Pass Filter Capacitance($C_3$=$C_4$) | 150 pf |
| 8 | Capacitor $C_1$ | 50 pf |
| 9 | Capacitor $C_2$ | 500nf |
| 10 | Local carrier amplitude | 450 mV |
| 11 | Local carrier frequency fc | 2 MHz |
| 12 | Low Pass filter cut off frequency | 60 KHz |
| 13 | Low Pass Resistance R2 | 200 Ω |
| 14 | Low Pass Filter capacitance ($C_1$=$C_3$) | 10pf |
| 15 | Bias voltage $V_{B1,3}$ | 600mv |

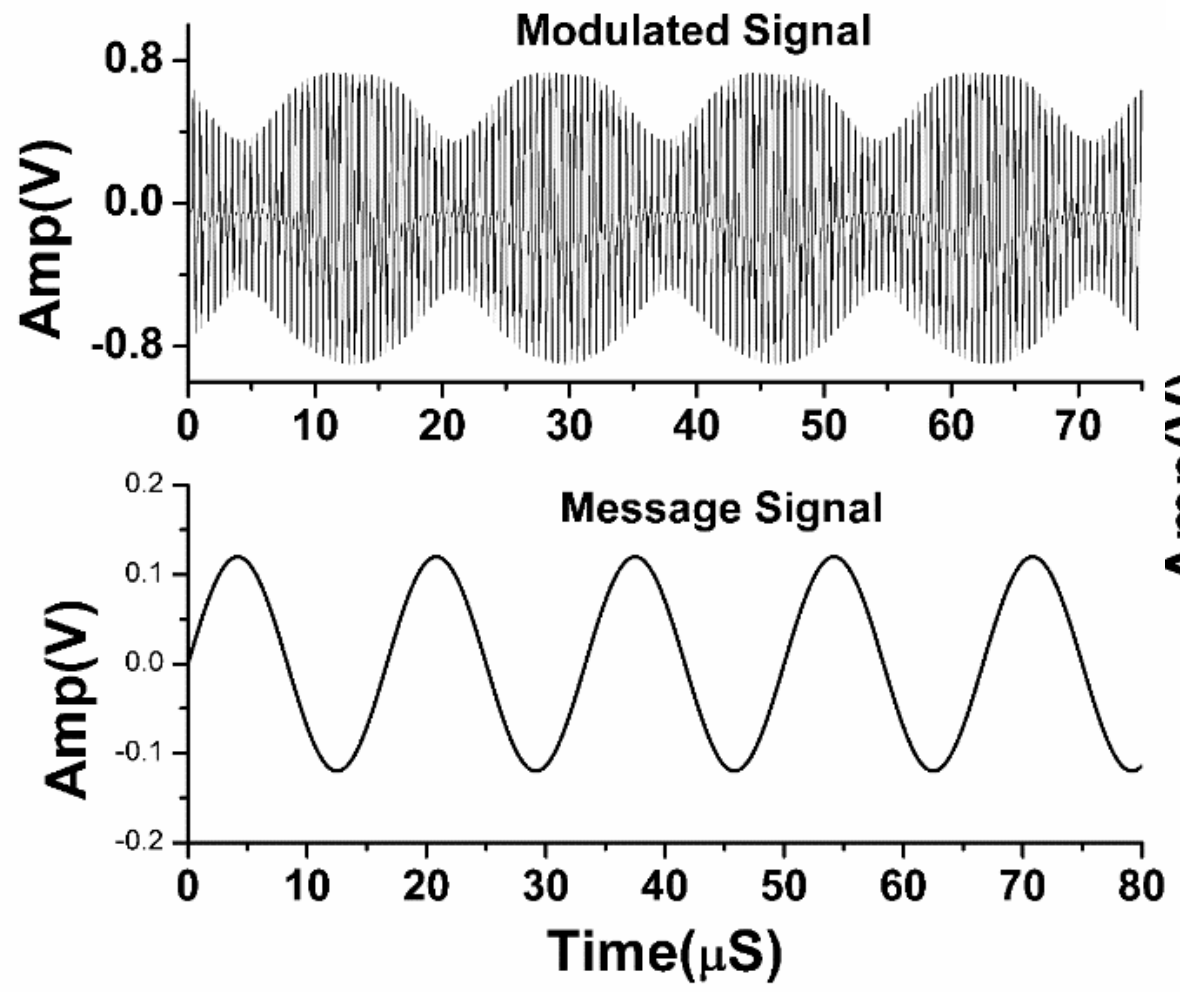

(a)

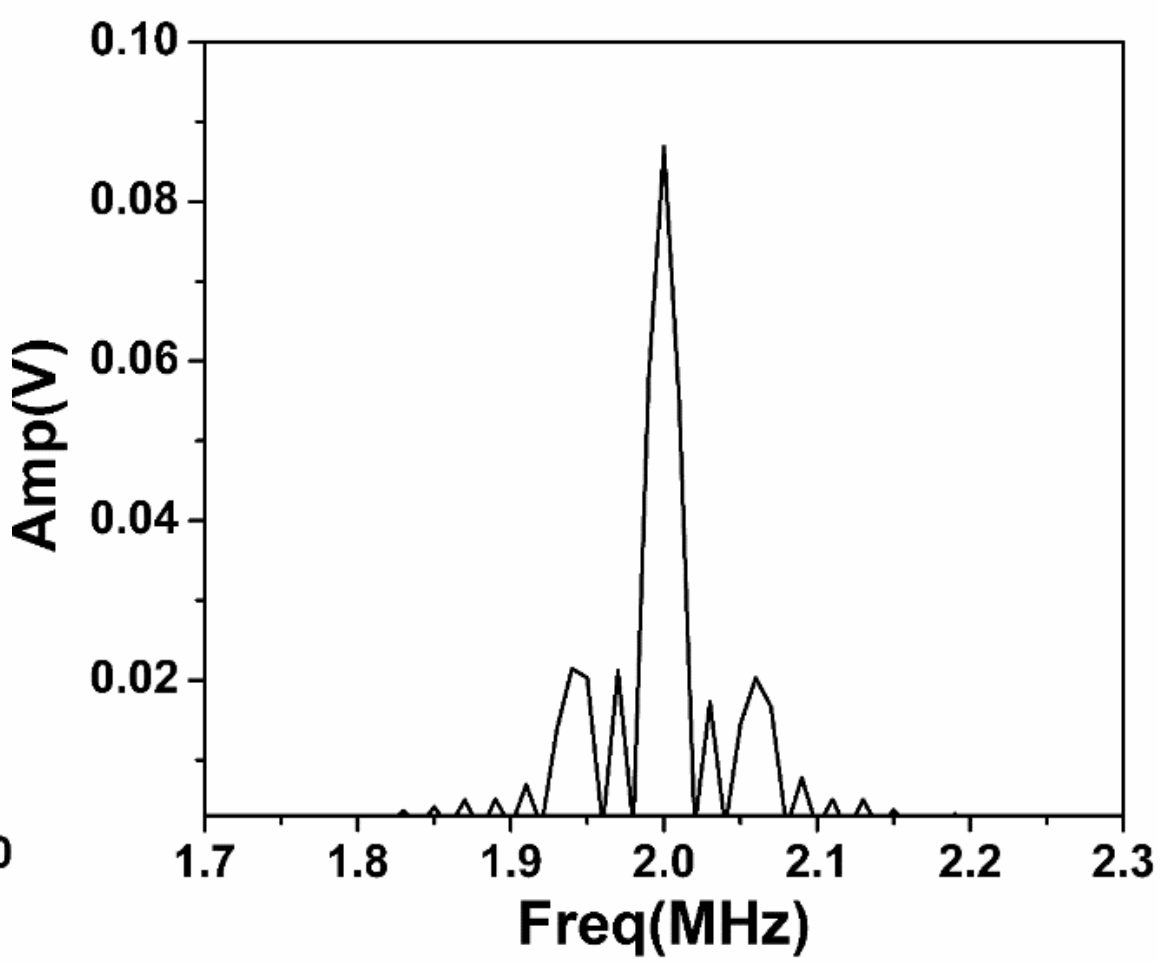

(b)

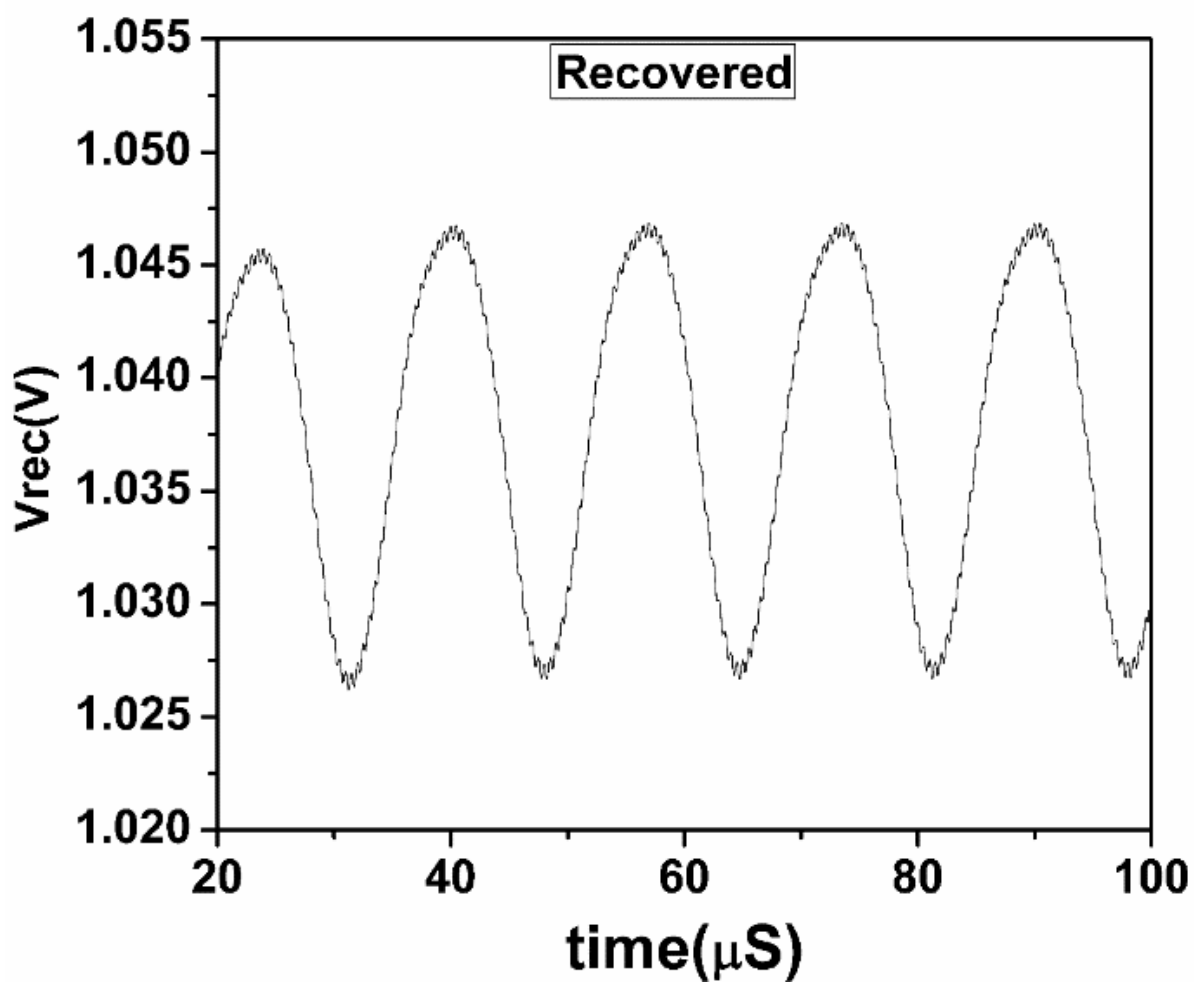

(c)

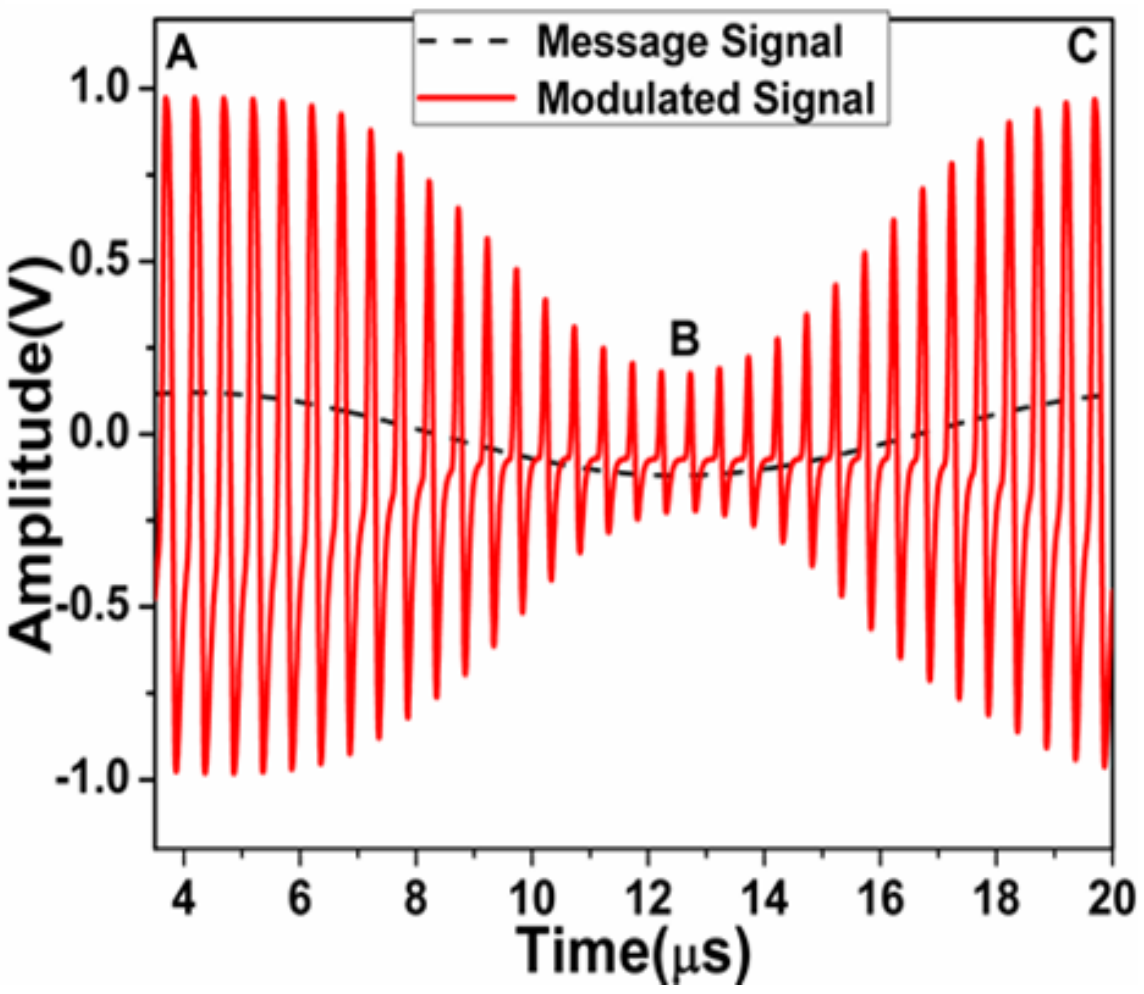

(d)

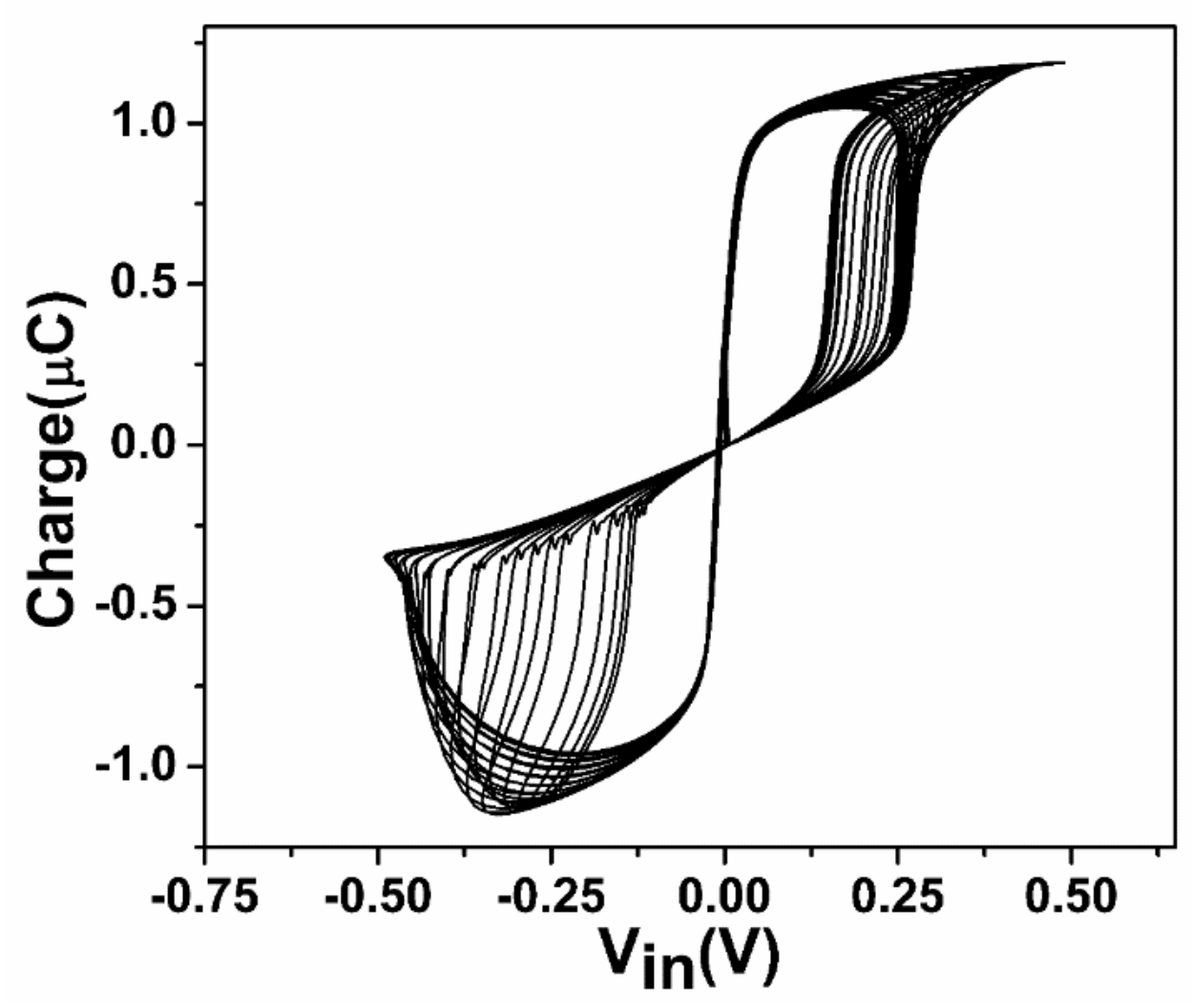

(e)

**Fig.20.** Plot of (a) message signal and modulated signal, (b) spectrum of the modulated signal, (c) demodulated signal, (d) waveform of message and carrier for one-time period, (e) hysteresis loop of AM for one-time period

## *8.4 Low pass filter using Memcapacitor*

To further verify the proposed circuit, the simplest possible implementation as RC low pass filter is shown in Fig .21(a). At low frequencies, capacitor acts as an open circuit and hence get voltage at the output terminal and at high frequencies capacitor acts as a short circuit and hence unable to get voltage at the output terminal. This circuit is implemented when Capacitor is replaced by Memcapacitor. Fig 21(b) shows the frequency response at the memcapacitance output of LPF.

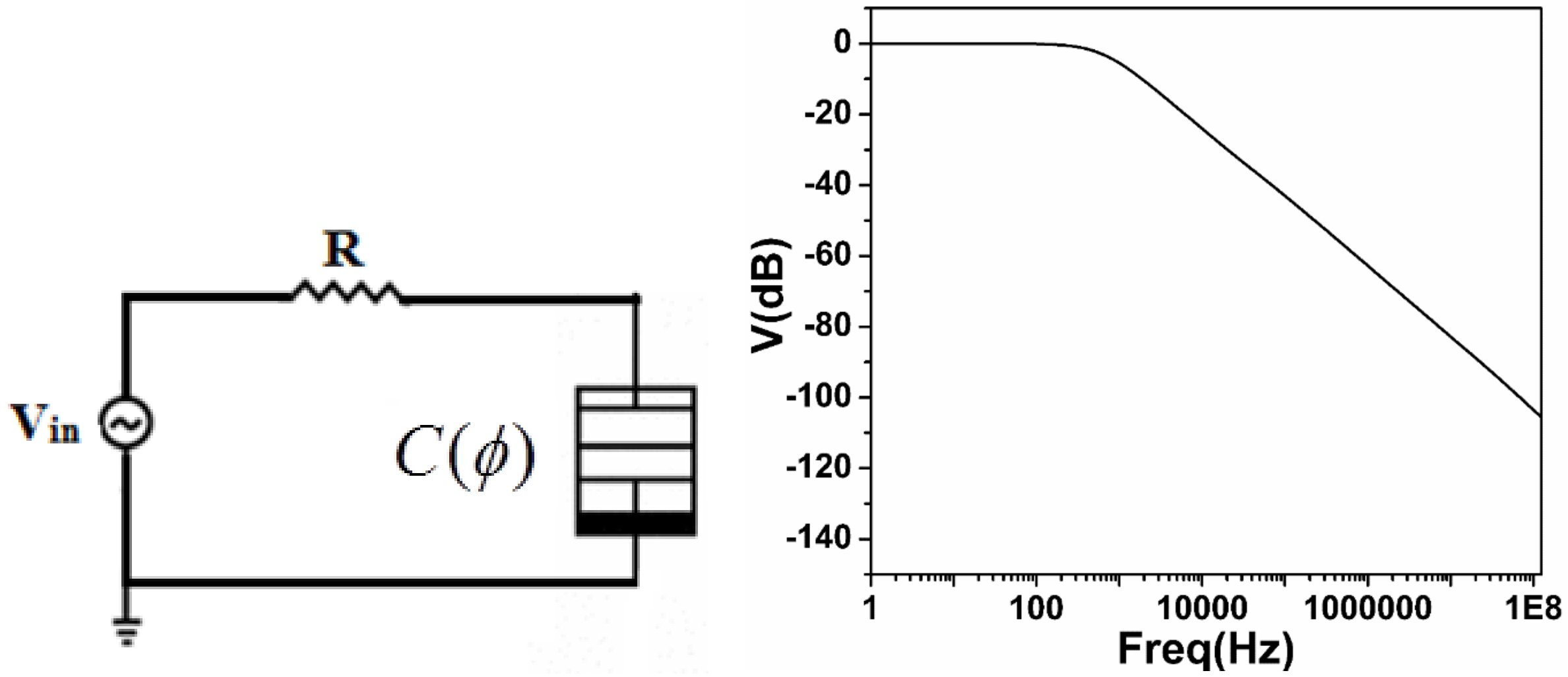

**(a)** **(b)**

**Fig 21.** (a) Memcapacitor circuit as $RC_M$ low pass circuit, (b) Frequency response of $RC_M$ low pass filter.

### *8.5 Memcapacitor inductor oscillator*

For small signal, the memcapacitance can be expressed as in [19]

$$C(\phi) = C_0 + K\phi(\mathrm{t})$$

For the initial energy stored in the system of Fig 22(a) the system starts oscillating. Such a circuit can be thought as a nonlinear oscillation circuit and the differential equation describing it can be obtained using Kirchhoff's laws and the components terminal equations. On routine analysis, we get the differential equation describing the non-linear oscillation as

$$\mathrm{LC}_0 \frac{d^2 V_C}{dt^2} + V_C + LK\phi(t)V_C = 0 \quad (35)$$

(35) is a non-linear equation due to term $\phi(t)$ , $V_C(t)$ and therefore lacks an analytic solution. Further analysis of the circuit to find the approximate solution using Perturbation theory results in

$$V_C(t) = \frac{d\phi}{dt} = V_o(t) + \varepsilon V_1(t) + \varepsilon^2 V_o(t) + ...... + \varepsilon^n V_n(t) \quad (36)$$

and

$$\mathrm{i_L(t)} = -\mathrm{KV}^2 + \left(\overline{C} + K(ACos(\omega_0\,\mathrm{t})) - \frac{KLA^2}{3}Cos(2\omega_0\,\mathrm{t})\right)\left(-\omega_0{}^2 ACos(\omega_0\,\mathrm{t}) + \frac{KLA^2}{3}4\omega_0{}^2 Cos(\omega_0\,\mathrm{t})\right) \quad (37)$$

where $V_n(t)$ is the nth order solution, A is the constant calculated from the initial condition and $\overline{C}$ is the average flux. The simulations are performed for memcapacitor components $C_1$=200pf, $C_2$=500nf and L=1nH. Fig.22(b-c) shows the current across the inductor and voltage across the memcapacitor respectively. Fig. 22(d) shows the plot of memcapacitor voltage with respect to current. It shows that the memcapacitor voltage vs. the inductor current is not an ellipse due to the circuit's nonlinearity as predicted by (36) and (37).

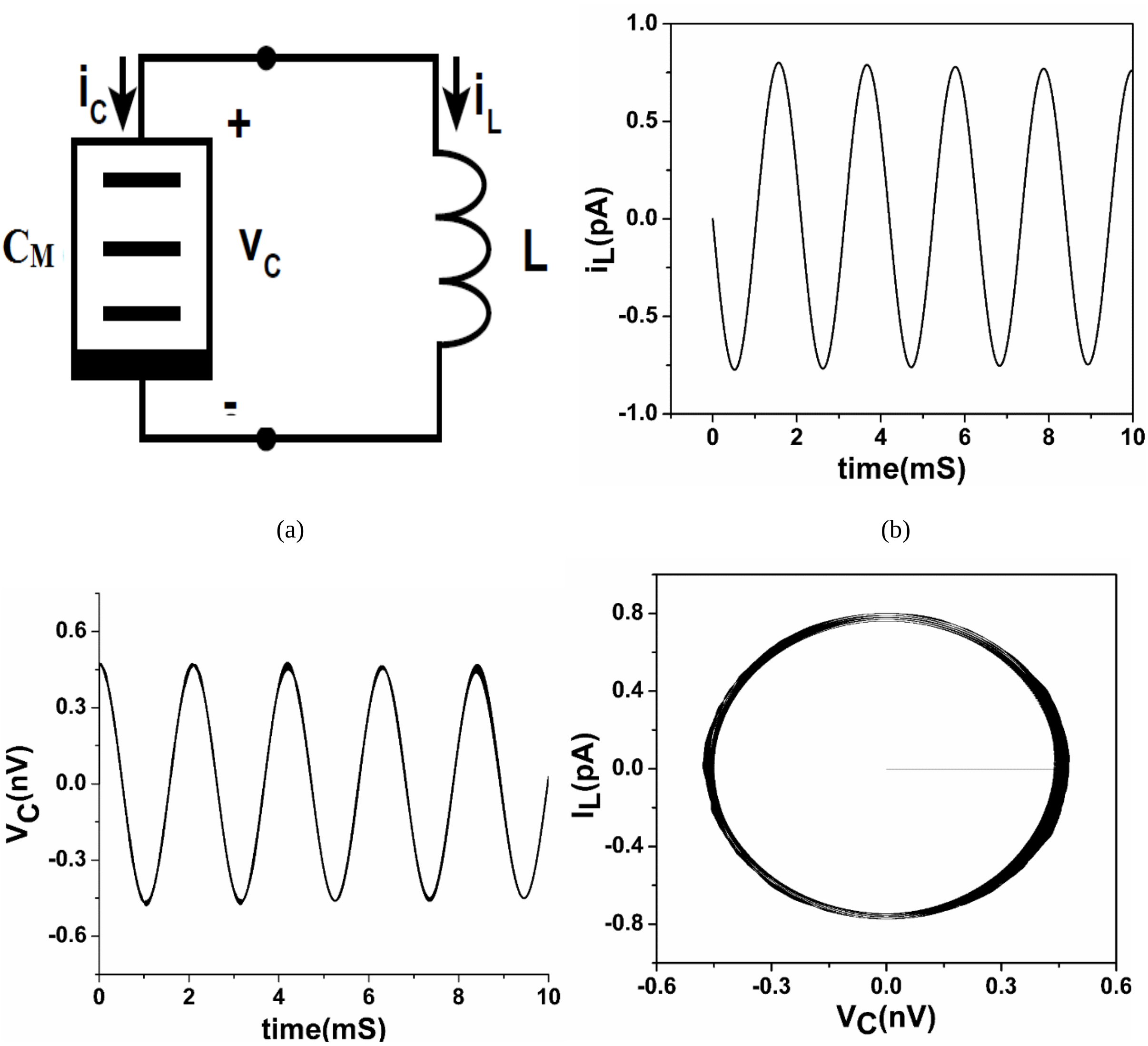

(a) (b)

(c) (d)

Fig.22 (a) Memcapacitor -Inductor Oscillator circuit, (b) Inductor current with respect to time, (c) Memcapacitor voltage with respect to time, (d) Memcapacitor voltage with respect to its current

## *8.6 Point attractor and periodic doubler*

Fig. 23(a) shows the non-linear oscillator circuit using memcapacitor [32]. Point attractors between ($\sigma$ - $i_{L3}$) and ($\sigma$ – $i_{L4}$) are shown in Fig. 23(b) and Fig. 23(c) respectively. Phase portrait of periodic doubling oscillations between $i_{L3}$ and $i_{L4}$ are shown in Fig 23(d). Simulations are performed for $L_3$= 465mH, $L_4$= 530mH, G= 600Ω, R=1.8 kΩ and $V_{B1,3}$= 500mV.

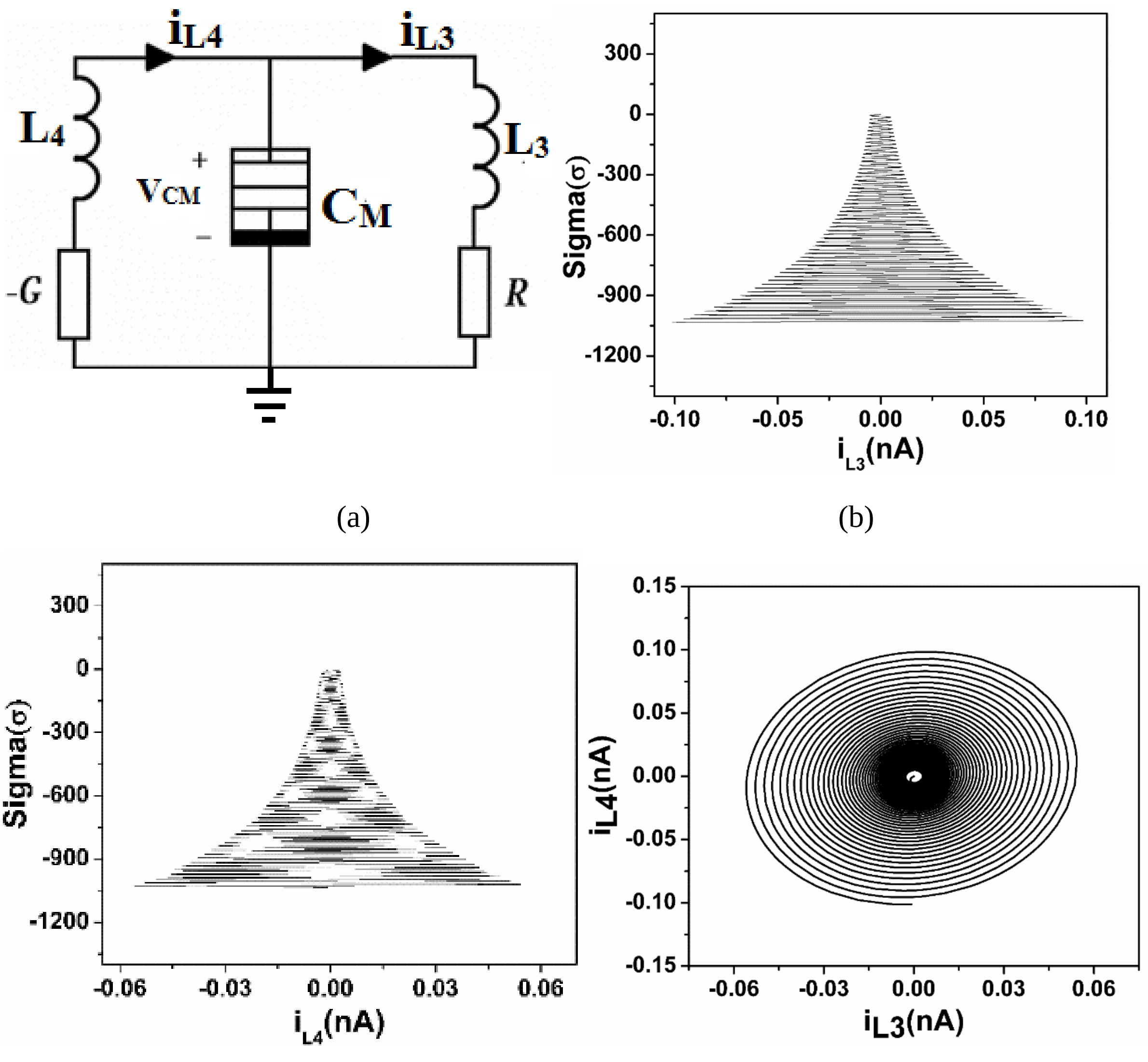

Fig.23 (a) Circuit for the chaotic oscillator, (b)point attraction between ($\sigma$- $i_{L3}$), (c) point attraction between ($\sigma$- $i_{L4}$), (d) Periodic doubling oscillation between ($i_{L3}$ - $i_{L4}$)

## 9. Experimental results

To verify experimentally, the proposed memcapacitor emulator is implemented as shown in Fig. 24(a) with capacitance(C1) as 30nf (in parallel combination of three 10nf) and commercially available BJT based OTA ICs (CA3080) and CFOF IC AD844. The prototype of the circuit is constructed on a breadboard as shown in Fig.24(b). The pinched hysteresis loops obtained for the operating frequency of 910 KHz for a 5V peak to peak input signal is shown in Fig. 24(c). However, because of limitations of operating bandwidth of 2MHz for IC CA3080 as well as parasitic faced due to interconnect of the experimental setup, the operating frequency range of the memristor emulator circuit on a breadboard is found to be close to 1 MHz. Fig. 24(d) shows the time domain waveform for charge and sigma (integration of charge). Fig. 24(e) shows the time domain waveform for input voltage and phi (integration of voltage). It can further be noted that since input to the circuit is a sinusoidal signal so the sigma and phi curve tends to decrease as time increase. This results in the overall increase in graph between sigma and phi. It can further be noted that the integration of the input signal and charge across the capacitor is performed through inbuilt integration function in oscilloscope to prevent and kind of loading effect and loss in signal.

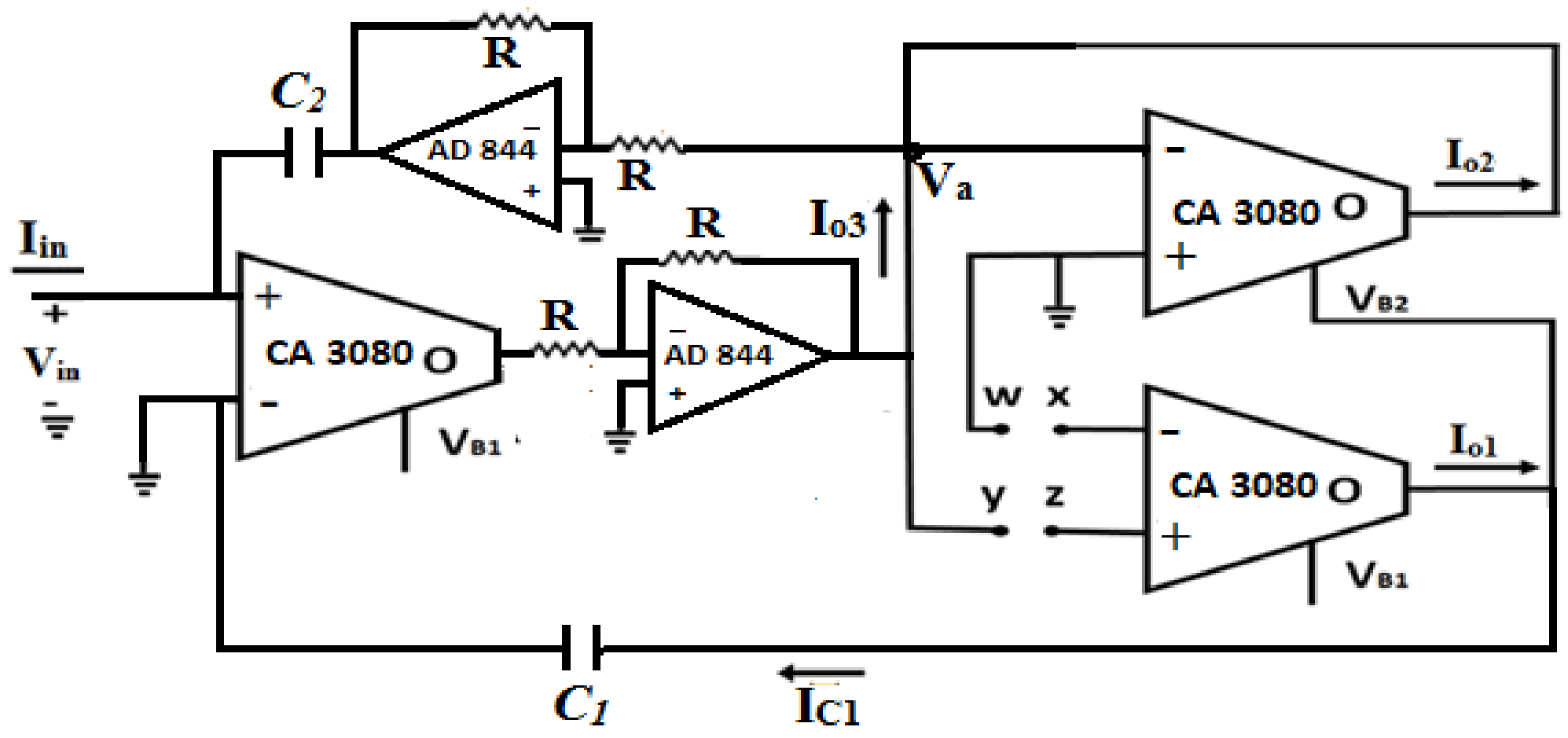

(a)

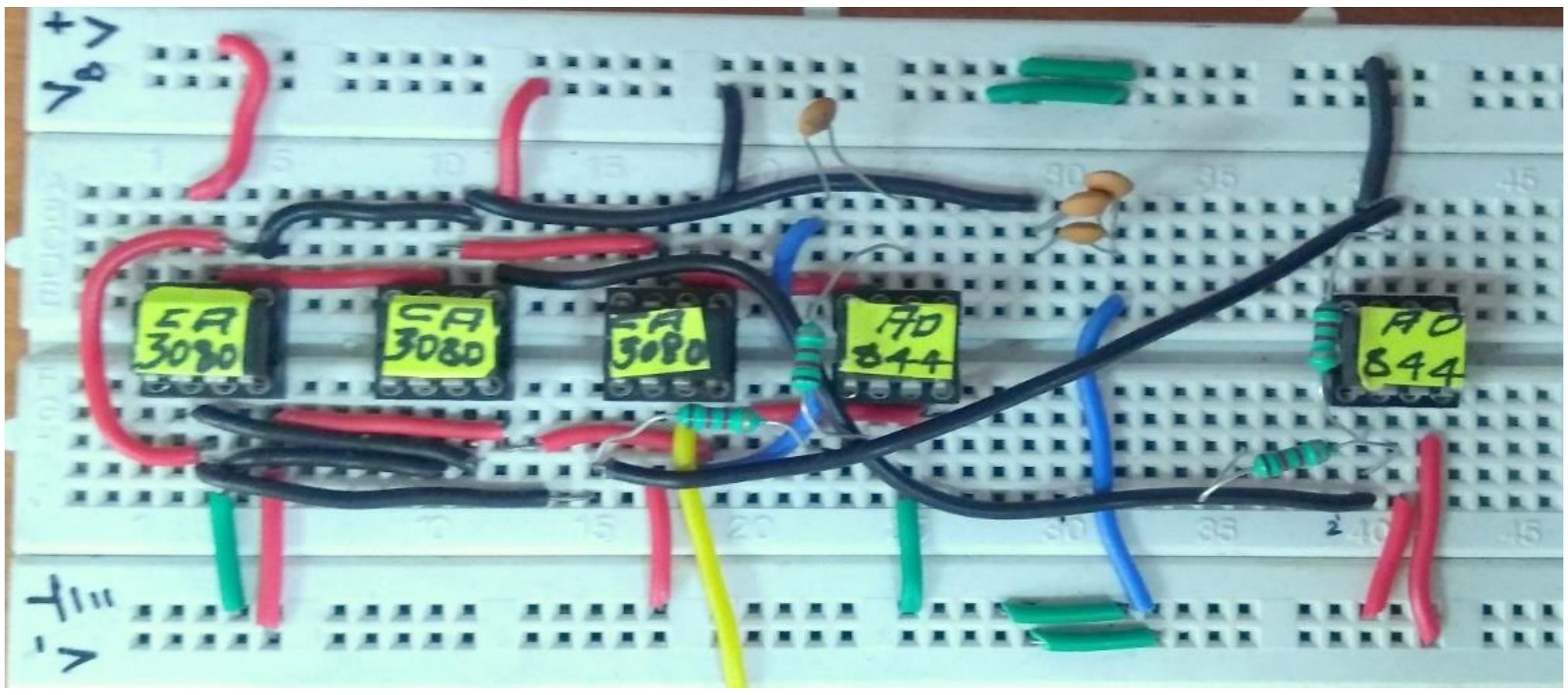

(b)

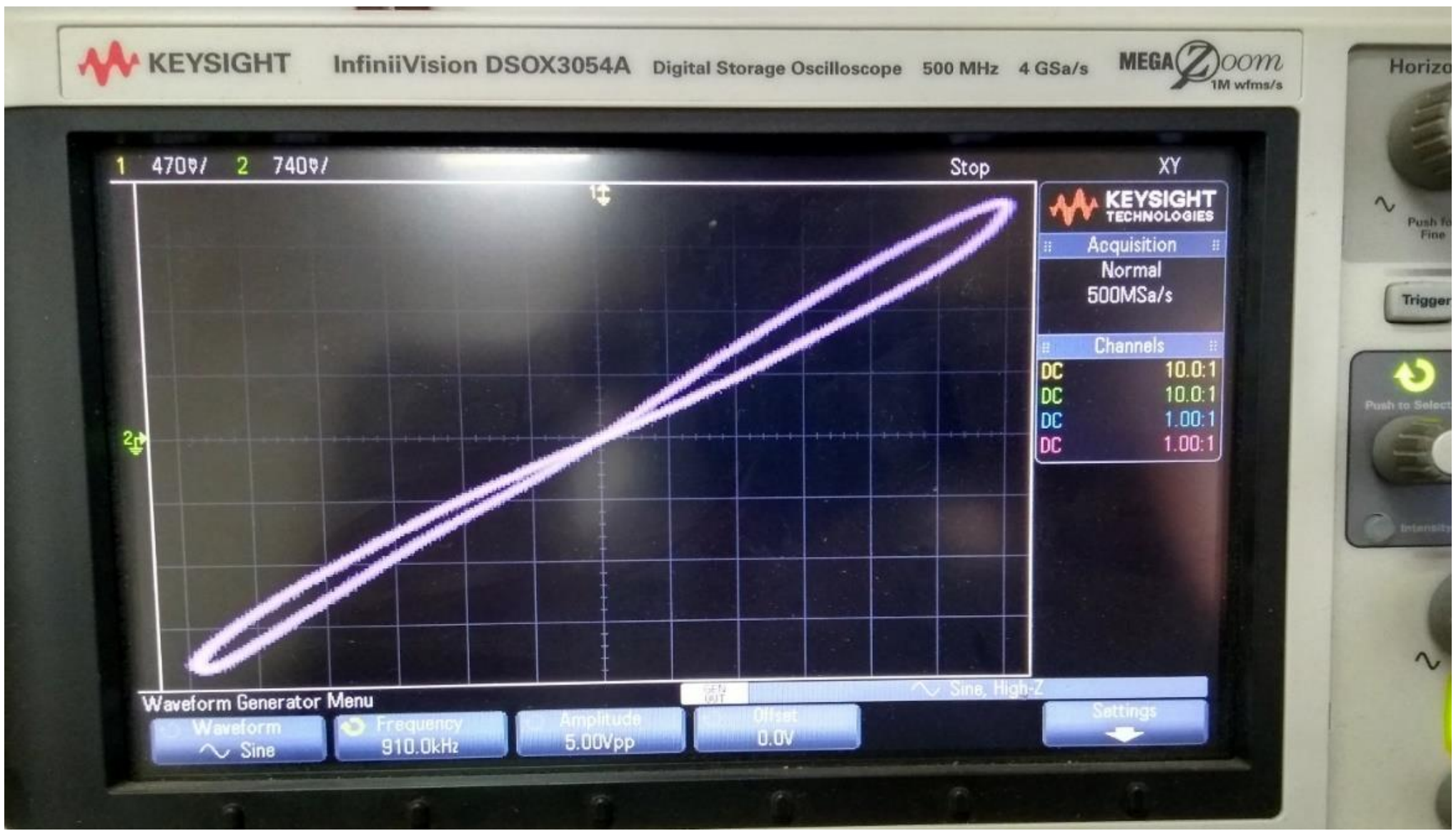

(c)

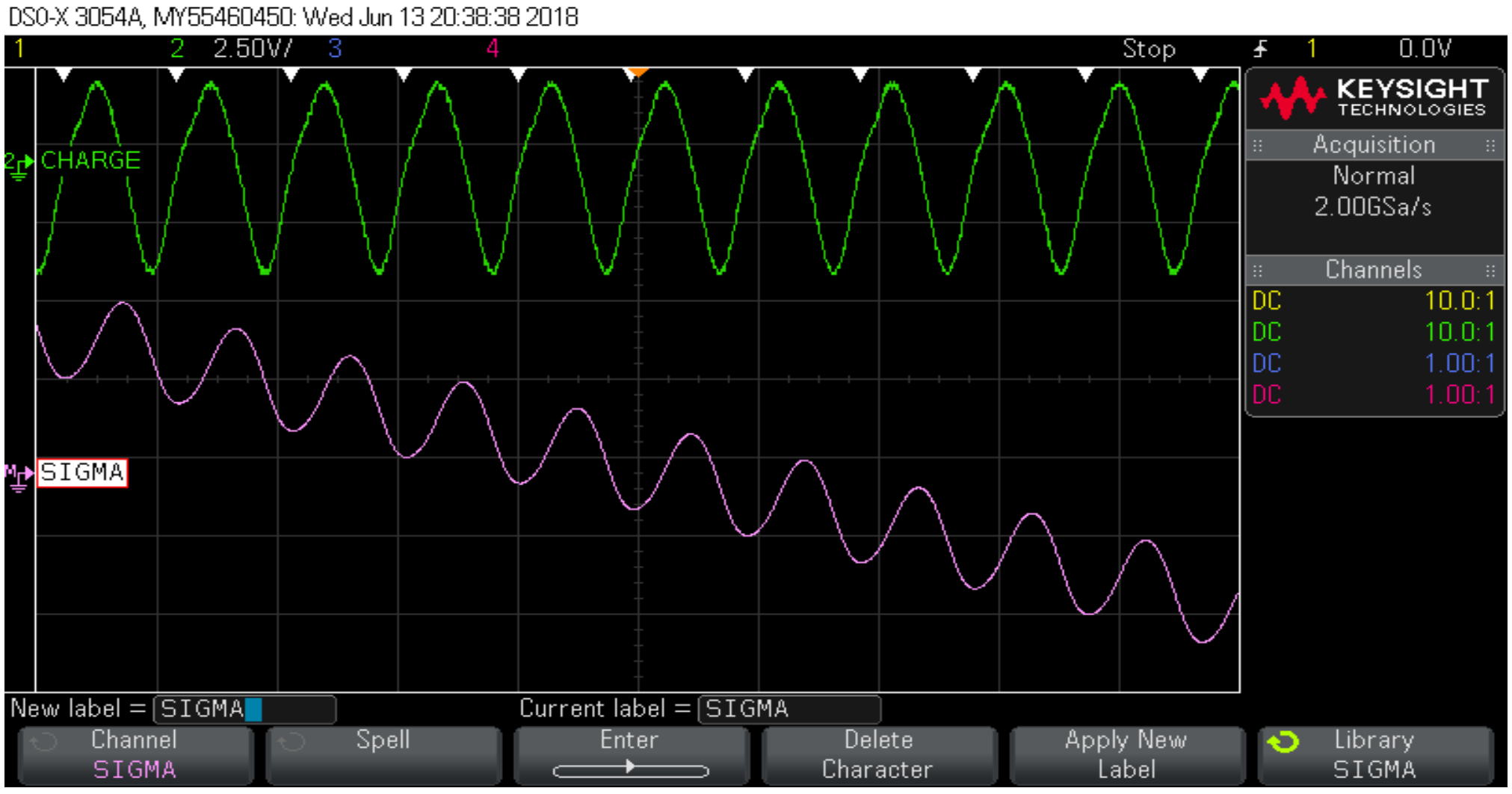
DSO-X 3054A, MY55460450: Wed Jun 13 20:38:38 2018
2 2.50V/
Stop
1 0.0V
CHARGE
SIGMA
KEYSIGHT TECHNOLOGIES
Acquisition
Normal
2.00GSa/s
Channels
DC 10.0:1
DC 10.0:1
DC 1.00:1
DC 1.00:1
New label = SIGMA
Current label = SIGMA
Channel SIGMA
Spell
Enter
Delete Character
Apply New Label
Library SIGMA

(d)

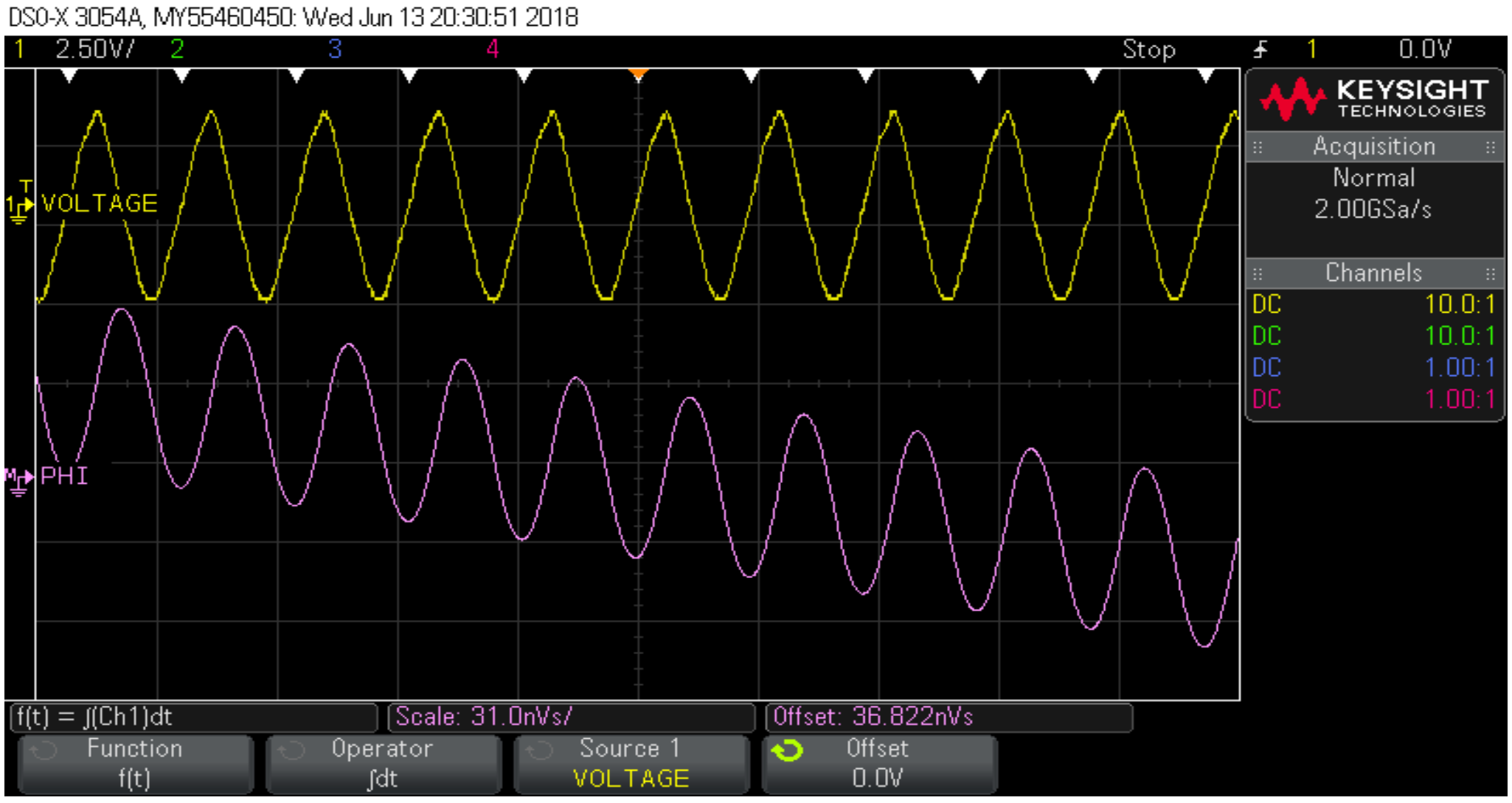

(e)

Fig. 24. Memcapacitor emulator (a) Experimental circuit, (b) Prototype on breadboard, (c) Experimental hysteresis loop, (d) Time domain waveform for Charge and Sigma waveform, (e) Time domain waveform for Voltage and phi

Fig. 25(a) represents the experimental setup for application of memcapacitor emulator as $RC_M$ low pass filter. Table 7 shows the experimental gain obtained for different input signal frequency at $C_1$= 40 nf, $C_2$= 10nf, $R_1$=3.6KΩ and input signal amplitude as 870 mV. Fig 25(b) shows that plot of frequency vs. gain for data obtained in Table 7. It shows that the proposed circuit works as a low pass filter.

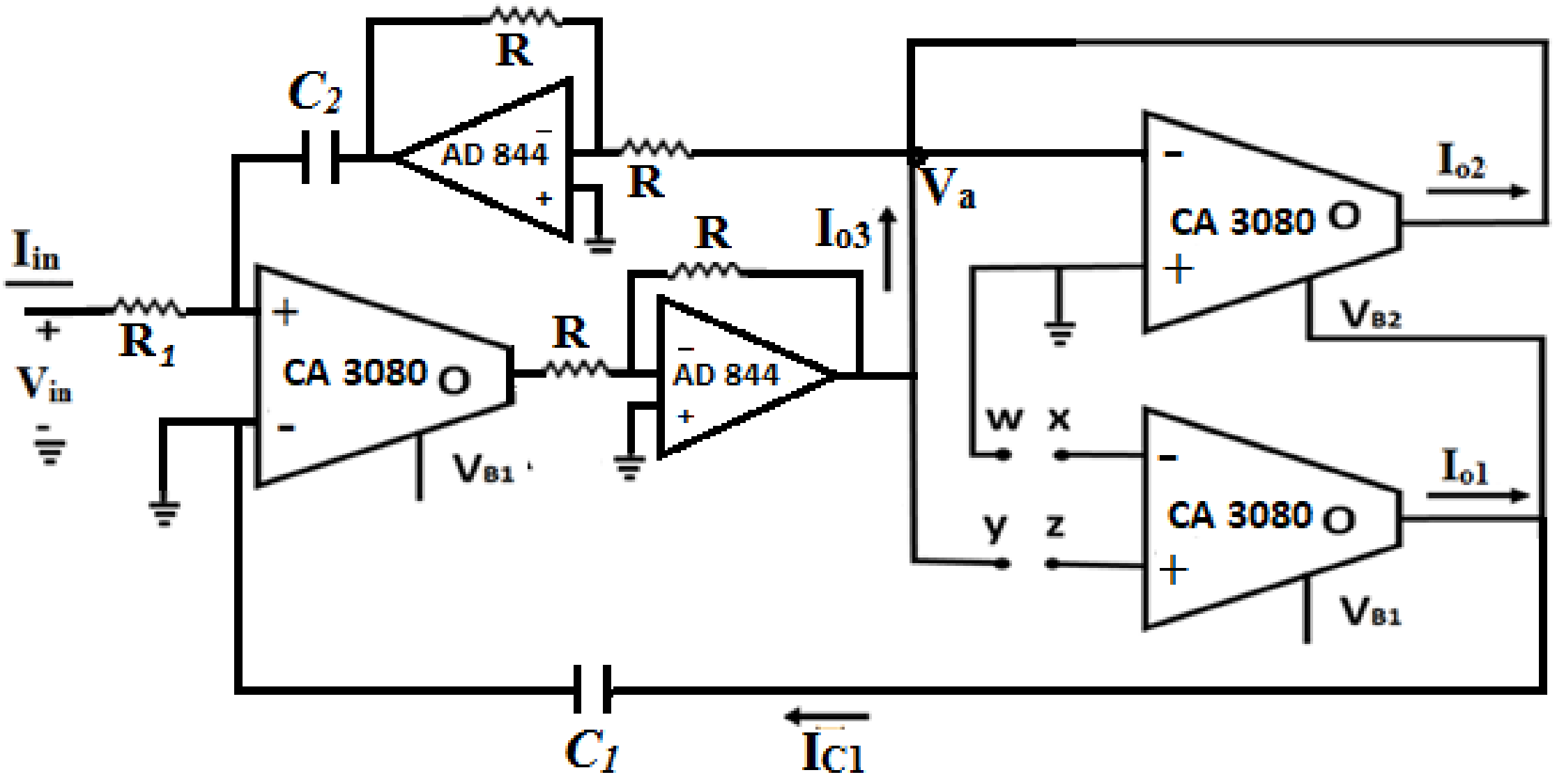

(a)

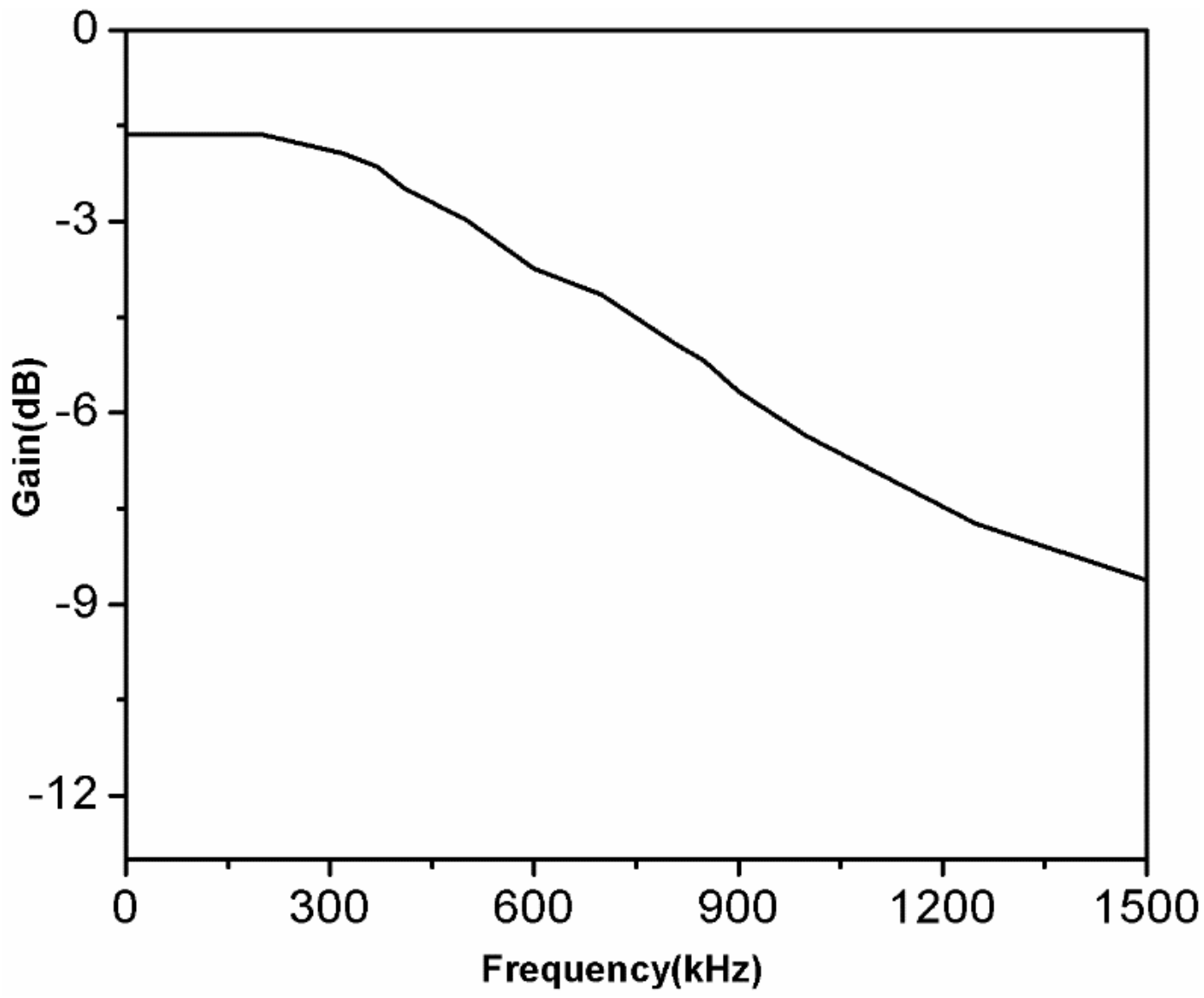

(b)

Fig. 25. (a) Experimental circuit of memcapacitor as $RC_M$ low pass circuit, (b) Gain vs. frequency plot for circuit in Fig. 25(a)

**Table 7.** Experimental frequency vs gain data

| Frequency (KHz) | \|GAIN(dB)\| |
|---|---|
| 0.5 | 1.649 |
| 200 | 1.649 |
| 320 | 1.938 |
| 410 | 2.498 |
| 500 | 2.974 |
| 600 | 3.741 |
| 700 | 4.152 |
| 800 | 4.882 |
| 900 | 5.679 |
| 1000 | 6.375 |
| 1500 | 8.635 |

# 10.Conclusion

The proposed memcapacitive emulators show pinched hysteresis loop similar to a real memcapacitive device. They have a simple circuitry built with only three OTAs and two capacitors. These emulators are first of their kind having high frequency of operation and resistor less topology. Proposed emulator circuits have been tested for a frequency range up to 8MHz in both incremental and decremental topology. Memcapacitance is electronically tuneable by the bias voltage of OTAs. Controllability of the pinched hysteresis loop and memcapacitance nature of proposed emulators for different frequency of input signal, passive components, and external bias voltages is verified by simulation results and has been discussed. Layout, post-layout simulation, detailed non-ideal analysis, and Monte Carlo results are also given. Moreover, as an application, an AM modulator has been realized using the proposed memcapacitor emulator circuit and verified its performance by simulation. Further applications such as filter and oscillators are also discussed.

## References

[1]. Chua LO. Memristor- the missing circuit element. IEEE Transaction on Circuit Theory 1971; 18(5):507-9.

[2]. Chua L. Device modeling via nonlinear circuit elements. IEEE Transactions on Circuits and Systems. 1980 Nov;27(11):1014-44.

[3]. Di Ventra M, Pershin YV, Chua LO. Circuit elements with memory: memristors, memcapacitors, and meminductors. Proceedings of the IEEE. 2009 Oct;97(10):1717-24.

[4]. Yin Z, Tian H, Chen G, Chua LO. What are memristor, memcapacitor, and meminductor?. IEEE Transactions on Circuits and Systems II: Express Briefs. 2015 Apr;62(4):402-6.

[5]. Biolek. D, Biolek. Z, Biolkova.V (2011). Pinched hysteretic loops of ideal memristors, memcapacitors and meminductors must be 'self-crossing'. Electronics Letters. 47. 1385-1387.

[6]. Pershin YV, Di Ventra M. Memristive circuits simulate memcapacitors and meminductors. Electronics Letters. 2010 Apr 1;46(7):517-8.

[7]. Biolek D, Biolková V, Kolka Z. Mutators simulating memcapacitors and meminductors. InCircuits and Systems (APCCAS), 2010 IEEE Asia Pacific Conference on 2010 Dec 6 (pp. 800-803). IEEE.

[8]. Biolek D, Biolkova V. Mutator for transforming memristor into memcapacitor. Electronics letters. 2010 Oct;46(21):1428-9.

[9]. Pershin YV, Di Ventra M. Emulation of floating memcapacitors and meminductors using current conveyors. Electronics Letters. 2011 Feb 17;47(4):243-4.

[10]. Fouda ME, Radwan AG. Charge controlled memristor-less memcapacitor emulator. Electronics letters. 2012 Nov 8;48(23):1454-5.

[11]. Wang XY, Fitch AL, Iu HH, Qi WG. Design of a memcapacitor emulator based on a memristor. Physics Letters A. 2012 Jan 9;376(4):394-9.

[12]. Yu DS, Liang Y, Chen H, Iu HH. Design of a practical memcapacitor emulator without grounded restriction. IEEE Transactions on Circuits and Systems II: Express Briefs. 2013 Apr;60(4):207-11.

[13]. Sah MP, Budhathoki RK, Yang C, Kim H. Expandable circuits of mutator-based memcapacitor emulator. International Journal of Bifurcation and Chaos. 2013 May;23(05):1330017.

[14]. Yu D, Liang Y, Iu HH, Chua LO. A universal mutator for transformations among memristor, memcapacitor, and meminductor. IEEE Transactions on Circuits and Systems II: Express Briefs. 2014 Oct;61(10):758-62.

[15]. Flak J, Lehtonen E, Laiho M, Rantala A, Prunnila M, Haatainen T. Solid-state memcapacitive device based on memristive switch. Semiconductor Science and Technology. 2014 Sep 18;29(10):104012.

[16]. Biolek D, Biolková V, Kolka Z, Dobeš J. Analog emulator of genuinely floating memcapacitor with piecewise-linear constitutive relation. Circuits, Systems, and Signal Processing. 2016 Jan 1;35(1):43-62.

[17]. Fouda ME, Radwan AG. Resistive-less memcapacitor-based relaxation oscillator. International Journal of Circuit Theory and Applications. 2015 Jul 1;43(7):959-65.

[18]. Yu D, Zhou Z, Iu HH, Fernando T, Hu Y. A coupled memcapacitor emulator-based relaxation oscillator. IEEE Transactions on Circuits and Systems II: Express Briefs. 2016 Dec;63(12):1101-5.

[19]. Yener ŞÇ, Mutlu R. Small signal model of memcapacitor-inductor oscillation circuit. InElectric Electronics, Computer Science, Biomedical Engineerings' Meeting (EBBT), 2017 2017 Apr 20 (pp. 1-4). IEEE.

[20]. Arora A, Niranjan V. Low power filter design using memristor, meminductor and memcapacitor. InElectrical, Computer and Electronics (UPCON), 2017 4th IEEE Uttar Pradesh Section International Conference on 2017 Oct 26 (pp. 113-117). IEEE.

[21]. Li Y, Yang C, Yu Y, Díez FF. Research on low pass filter based on Memristor and memcapacitor. InControl Conference (CCC), 2017 36th Chinese 2017 Jul 26 (pp. 5110-5113). IEEE.

[22]. Noh YJ, Baek YJ, Hu Q, Kang CJ, Choi YJ, Lee HH, Yoon TS. Analog memristive and memcapacitive characteristics of Pt-Fe 2 O 3 core-shell nanoparticles assembly on p+-Si substrate. IEEE Transactions on Nanotechnology. 2015 Sep;14(5):798-805.

[23]. Pershin YV, Di Ventra M. Neuromorphic, digital, and quantum computation with memory circuit elements. Proceedings of the IEEE. 2012 Jun;100(6):2071-80.

[24]. Kim KH, Gaba S, Wheeler D, Cruz-Albrecht JM, Hussain T, Srinivasa N, Lu W. A functional hybrid memristor crossbar-array/CMOS system for data storage and neuromorphic applications. Nano letters. 2011 Dec 9;12(1):389-95.

[25]. Li C, Li C, Huang T, Wang H. Synaptic memcapacitor bridge synapses. Neurocomputing. 2013 Dec 25;122:370-4.

[26]. Fouda ME, Radwan AG. On the mathematical modeling of memcapacitor bridge synapses. InMicroelectronics (ICM), 2014 26th International Conference on 2014 Dec 14 (pp. 172-175). IEEE.

[27]. Hu Z, Li Y, Jia L, Yu J. Chaotic oscillator based on voltage-controlled memcapacitor. InCommunications, Circuits and Systems (ICCCAS), 2010 International Conference on 2010 Jul 28 (pp. 824-827). IEEE.

[28]. Hu Z, Li Y, Jia L, Yu J. Chaos in a charge-controlled memcapacitor circuit. InCommunications, Circuits and Systems (ICCCAS), 2010 International Conference on 2010 Jul 28 (pp. 828-831). IEEE.

[29]. Fitch AL, Iu HH, Yu DS. Chaos in a memcapacitor based circuit. InCircuits and Systems (ISCAS), 2014 IEEE International Symposium on 2014 Jun 1 (pp. 482-485). IEEE.

[30]. Wang G, Jiang S, Wang X, Shen Y, Yuan F. A novel memcapacitor model and its application for generating chaos. Mathematical Problems in Engineering. 2016;2016.

[31]. Yuan F, Wang G, Wang X. Chaotic oscillator containing memcapacitor and meminductor and its dimensionality reduction analysis. Chaos: An Interdisciplinary Journal of Nonlinear Science. 2017 Mar;27(3):033103.

[32]. Yuan F, Wang G, Shen Y, Wang X. Coexisting attractors in a memcapacitor-based chaotic oscillator. Nonlinear Dynamics. 2016 Oct 1;86(1):37-50.

[33]. Parveen, Tahira. Textbook of Operational Transconductance Amplifier and Analog Integrated Circuits. IK International Pvt Ltd, 2013. (5-6)

[34]. Whitaker JC, editor. The electronics handbook. CRC Press; 1996 Dec 23(666-669)

[35]. Deliyannis T, Sun Y, Fidler JK. Continuous-time active filter design. CRC Press; 1998 Dec 16 (241-244)

[36]. Al-Hashimi, B. Current mode filter structure based on dual output transconductance amplifiers. Electronics Letters, 1996; 32(1): 25-26.